%% file: main.tex
\documentclass[]{eptcs}

\usepackage[greek,english]{babel} 
\usepackage{makeidx}
\usepackage[nomargin,inline,index]{fixme} 

\FXRegisterAuthor{fxJL}{anfxJL}{\color{cyan} {\underline{Julien}}}
\FXRegisterAuthor{fxAS}{anfxAS}{\color{blue} {\underline{Alceste}}}

\input{styles.tex}
\input{macros.tex}
\input{jlmacro.tex}
\input{co2-macros.tex}

\title{Choreography Synthesis as Contract Agreement}

\author{
  Julien Lange
  \institute{University of Leicester, UK}
  \email{jlange@le.ac.uk}
  \and
  Alceste Scalas
  \institute{University of Cagliari, Italy}
  \email{alceste.scalas@unica.it}
}
 
\def\titlerunning{Choreography Synthesis as Contract Agreement}
\def\authorrunning{J.\ Lange and A.\ Scalas}

\begin{document}

\renewcommand{\emptyset}{\varnothing}

\maketitle


\begin{abstract}
  We propose a formal model for distributed systems, where
  each participant advertises its requirements and obligations as behavioural
  \emph{contracts}, and where multiparty sessions are started when
  a set of contracts allows to synthesise a 
  \emph{choreography}.
  Our framework is based on the \coco calculus for contract-oriented
  computing, and borrows concepts and results from the session type
  literature.
  %
  %
  
  It supports sessions where the number of participants is
  not determined beforehand, and keeps \coco's ability to rule out
  participants that are \emph{culpable} if contracts are not
  fulfilled at runtime.
  %
  We show that we have \emph{progress} and \emph{session fidelity}
  in \coco, as a result of the \emph{honesty} of participants ---
  i.e., their ability to always adhere to their contracts.
  %
\end{abstract}

\input{intro.tex}
\input{session.tex}

\input{co2.tex}
\input{result.tex}
\input{conc.tex}

%
\label{sect:bib}
\bibliographystyle{eptcsini}
\bibliography{co2,julien}

\end{document}

%% file: styles.tex
\usepackage{latexsym}
\usepackage{graphicx}
\usepackage{amssymb}
\usepackage{amsmath}
\usepackage{amssymb}
\usepackage{amsthm}
\usepackage{mathtools} 
\usepackage{nicefrac}  
\usepackage{colonequals} 
\usepackage{color}
\usepackage[usenames,dvipsnames]{xcolor}
\usepackage{latexsym}
\usepackage{graphicx}
\usepackage{url}
\usepackage{mathpartir}
\usepackage{stmaryrd}
\usepackage{wasysym}
\usepackage{pifont}
\usepackage{xyling} 
\usepackage{amscd} 
\usepackage{array}
\usepackage{multirow}
\usepackage{fixltx2e} 
\usepackage{xspace} 
\usepackage{IEEEtrantools}          
\usepackage{enumerate} 

\usepackage{rotating}

\usepackage{textcomp} 

\numberwithin{equation}{section}

\newtheorem{theorem}{Theorem}[section]

\newtheorem{definition}[theorem]{Definition}
\newtheorem{example}[theorem]{Example}

\usepackage{proof}                  

%% file: macros.tex
\newcommand{\suchthat}{\colon}

\newcommand{\setenum}[1]{\left\{ #1 \right\}} 
\newcommand{\setcomp}[2]{\left\{ #1 \mid #2 \right\}} 

\newcommand{\fv}[1]{\operatorname{fv}(#1)}

\newcommand{\dom}[1]{\operatorname{dom}(#1)}

\newcommand{\img}[1]{\operatorname{img}(#1)}

\newcommand{\defeq}{\mathbin{\vcentcolon=}}

\newcommand{\bnfdef}{\coloncolonequals} 
\newcommand{\bnfsep}{\ \;\;\big|\;\;\ } 

\newcommand{\subst}[2]{\left\{\nicefrac{#2}{#1}\right\}}

\newcommand{\clc}[1]{{\color{blue}#1}} 
\newcommand{\clsy}[1]{{\color{ForestGreen}#1}} 

\newcommand{\ptpNames}{\mathbb{P}}
\newcommand{\ptpVars}{\mathcal{P}}
\newcommand{\ptp}[1]{\mathtt{#1}} 
\newcommand{\ptpc}[1]{\ptp{\uppercase{#1}}}
\newcommand{\ptpv}[1]{\ptp{\lowercase{#1}}}
\newcommand{\ptpcv}[1]{\underline{\ptp{\lowercase{#1}}}} 

\newcommand{\coco}{CO\textsubscript{2}\xspace}

\newcommand{\sesNames}{\mathbb{S}}
\newcommand{\sesVars}{\mathcal{S}}
\newcommand{\sesFont}[1]{#1} 
\newcommand{\sesc}[1]{\sesFont{\lowercase{#1}}}
\newcommand{\sesv}[1]{\sesFont{\lowercase{#1}}}

\newcommand{\msgsort}[1]{\mathsf{#1}} 
\newcommand{\atom}[1]{\msgsort{#1}}

\newcommand{\rulename}[1]{\textsc{\footnotesize[#1]}}
\newcommand{\smallrulename}[1]{\textsc{\scriptsize[#1]}}
\newcommand{\ltsrule}[3]{\inferrule*[left=\smallrulename{#1}\ ]{#2}{#3}}
\newcommand{\ltsruleS}[2]{\smallrulename{\textsc{#1}} \ \ \ #2}


%% file: jlmacro.tex
\newcommand{\emptyQ}{[]}

\newcommand{\QDEF}{(\ptp{A B})}
\newcommand{\QUEUEN}[1]{\clsy{\QDEF : #1}}
\newcommand{\QUEUE}[2]{\clsy{(\ptp{#1}) \!: \! #2}}

\newcommand{\BNFOR}{\;\; \mid \;\; }

\newcommand{\SYSV}{\clsy{T}} 
\newcommand{\PV}{\clc{c}} 
\newcommand{\SORTV}{\msgsort{e}} 
\newcommand{\GA}{\mathcal{G}} 

\newcommand{\GSENDPREF}[2]{\ptp{#1} \! \rightarrow \! \ptpc{#2} \!:\!}
\newcommand{\GSEND}[3]{\GSENDPREF{#1}{#2} \msgsort{#3}}

\DeclareMathOperator{\GCH}{+}
\DeclareMathOperator{\basePAR}{\; \mid \;}
\newcommand{\GPAR}{\basePAR}
\newcommand{\GEND}{\mathbf{0}}
\DeclareMathOperator{\GSEP}{;}

\newcommand{\GRECN}{\mu \, \PRECV{x} .}

\newcommand{\GRECV}{\PRECV{x}}

\newcommand{\PSEP}{;}

\newcommand{\PSEND}[2]{\clc{\ptp{#1}  !  \msgsort{#2}}}
\newcommand{\PRECEIVE}[2]{\clc{\ptp{#1}  ? \msgsort{#2}}}

\DeclareMathOperator{\PINCH}{\oplus}
\DeclareMathOperator{\POUTCH}{+}

\newcommand{\PSUM}[2]{\clc{\sum_{#1}#2}}
\newcommand{\PINS}[2]{\clc{\bigoplus_{#1}#2}}
\newcommand{\PPAR}{\clsy{\basePAR}}
\newcommand{\END}{\clsy{\mathbf{0}}}
\newcommand{\PEND}{\clc{\mathbf{0}}}
\newcommand{\PBOX}[2]{\clsy{\ptp{#1} \! \left\langle {\clc{#2}} \right\rangle}}

\newcommand{\PPROD}[2]{\big|_{#1} #2}
\newcommand{\PRECN}{\clc{\mu \, \PRECVN .}}

\newcommand{\PRECV}[1]{\mathbf{#1}}
\newcommand{\PRECVN}{\clc{\PRECV{x}}}

\newcommand{\labsend}[3]{#1 \rightarrow #3 : #2}
\newcommand{\labrec}[3]{#1 \leftarrow #3 : #2}

\newcommand{\semLab}[1]{\xrightarrow{#1}}
\newcommand{\semSend}[3]{\semLab{\labsend{#1}{\msgsort{#2}}{#3}}}
\newcommand{\semRecv}[3]{\semLab{\labrec{#1}{\msgsort{#2}}{#3}}}
\newcommand{\semSendOrRecv}[3]{\xrightarrow{\ptp{#1}
    \leftrightharpoons \ptp{#3} : \msgsort{#2}}}

\newcommand{\PARTS}[1]{\mathcal{P}(#1)} 
\newcommand{\QUEUES}[1]{\clsy{\operatorname{\textsc{q}}(#1)}} 

\newcommand{\emptyctx}{\circ}
\newcommand{\gloturn}{\vdash}
\newcommand{\glotri}{\blacktriangleright}

\newcommand{\judgeC}[3]
{#1  \gloturn #2 \, \glotri #3}
\newcommand{\judge}[3]
{\judgeC{#1}{#2}{#3}}



%% file: co2-macros.tex
\def\pmv{\ptpc}

\newcommand{\CSY}{S} 

\newcommand{\cocoAgreement}[4]{{#3} \vartriangleright^{#1}_{#2} {#4}}

\newcommand{\cocoAllReadysets}[1]{{\operatorname{CRS}}\!\left(#1\right)}

\newcommand{\cocoAllReadydo}[3]{\operatorname{PRS}_{\ptpc{#2}}^{#1}
  \!\left({#3}\right)}

\newcommand{\cocoAllWeakreadydo}[3]{\operatorname{WPRS}
  _{\ptpc{#2}}^{#1}\!\left({#3}\right)}

\newcommand{\cocoPassivearrow}[3]{\xrightarrow{\neq (\ptpc{#2} \colon
    \cocoDo{#1}{}{\!})}}

\newcommand{\cocoSays}{\;\mathit{says}\;}
\newcommand{\cocoSeq}{\mathbin{.}}
\newcommand{\cocoPlus}{\mathbin{+}}
\def\cocoSum{\sum}

\newcommand{\cocoMove}[1]{\xrightarrow{#1}}
\newcommand{\cocoCmove}[3]{\semSendOrRecv{#1}{#2}{#3}}

\newcommand{\cocoSys}[2]{{#1}\!\left[{#2}\right]}
\newcommand{\cocoEmptysys}{\mathbf 0}

\newcommand{\cocoTau}{\tau}
\newcommand{\cocoPar}{\mid}

\newcommand{\cocoFreeze}[2]{\downarrow_{#1}{#2}}

\newcommand{\cocoDo}[3]{\mathsf{do}^{#1}_{#2}\,{\atom{#3}}}
\newcommand{\cocoTell}[3]{\mathsf{tell}_{\ptpc{#1}}\!\cocoFreeze{#2}{\clc{#3}}}

\newcommand{\cocoFuse}{\mathsf{fuse}}

\newcommand{\cocoFuseN}[1]{\cocoFuse[#1]}
\newcommand{\cocoFuseR}{\cocoFuse_{R}}
\newcommand{\cocoFuseT}{\cocoFuse_{T}}

\newcommand{\cocoEmptyproc}{\mathbf 0}


%% file: intro.tex
\section{Introduction}\label{sec:intro}

Distributed applications are nowadays omnipresent but even for
seemingly simple cases, there is still a pressing need to make sure
they do work as their designers intended.
Indeed, such systems are difficult to design, verify, implement, deploy, and maintain.
Besides the intrinsic issues due to the underlying execution model
(concurrency, physical distribution, etc.), applications have to be designed
within a strange paradox: they are made of components that, on the one hand, collaborate with each
other and, on the other hand, may compete for resources, or for achieving conflicting goals.
This paradox is especially relevant in inter-organisational
service-oriented scenarios, where services may be deployed by
different entities:
even under common policies, the implementations may reflect diverging
and changing requirements, up to the point of departing from the agreed
specifications.
This issue is reflected in standards such as \cite{Oasis2012SOARef01},
which includes runtime monitoring and logging to check that
interactions in SOAs actually adhere to agreed policies and service
descriptions.

%
%

Along the lines of \cite{BZ10,BTZ12},
we propose a formal model for distributed systems where
\emph{contracts} drive interactions: components advertise behavioural
contracts; such contracts are used at runtime to establish multiparty
\emph{agreements}, and such agreements steer the behaviour of
components.
Therefore, contracts are not just a specification or a design
mechanism anymore, rather they become a pivotal element of the
execution model.

In this work, we combine two approaches:
\emph{session types}~\cite{hyc08} and \emph{contract-oriented
  computing}~\cite{BZ10}.
From the former, we adopt concepts, syntax and semantics --- and in particular,
the interplay between local behaviours and choreographies (i.e., between
local types and global types) as a method for specifying and analysing
the interactions of participants in a distributed system.
However,
in our framework we do not assume that a participant will necessarily
always adhere to its specification, nor that a global description is available
beforehand to validate the system.
From the contract-oriented computing approach, we adopt \coco~\cite{BTZ12},
a generic contract-oriented calculus
where participants advertise their requirements and obligations
through contracts, and interact with each other once
\emph{compliant} contracts have been found.
Here, we tailor \coco to a multiparty model where contracts have the
syntax of local types.
We say that contracts $c_1,\ldots,c_n$ are compliant when,
roughly, they can be used to synthesise a choreography ---
i.e., a global type whose projections are $c_1,\ldots,c_n$ themselves
\cite{LT12}.
Once a set of compliant contracts has been found, a \coco session may
be established, wherein the participants who advertised the contracts
can interact.
However, in line with what may happen in real life scenarios,
the runtime behaviour of these
participants may then depart from the contracts: the calculus allows
to model these situations, and reason about them.

\subsection{Contributions}
Our framework models multiparty contractual agreements as
``tangible'' objects, i.e., choreographies.
This allows us to rely on results and properties from the session type
literature --- in particular, the well-formedness of a choreography
ensures that contractual agreements enjoy knowledge of choice,
error/deadlock freedom, and progress.
Furthermore, it allows us to easily check that some meta-level properties are
satisfied at runtime, e.g., on the number of involved participants,
whether or not the session may terminate, etc.

Our adaptation of \coco to a multiparty, choreography-based contract
model preserves the properties of the original calculus.
In particular, if a system gets stuck, it is possible to identify
which participants violated their contracts.

We also discuss how the properties of a well-formed choreography are
reflected in a context where participants can misbehave.
We introduce global progress and session fidelity in \coco, again inspired by
 analogous concepts in theories based on session types.
We show that they hold in systems where all participants are
\emph{honest} (i.e., they always respect their contracts in any context)
--- even  when a participant takes part in
multiple sessions.

\paragraph{Synopsis.}
The rest of the paper is structured as follows.
In the rest of this section, we introduce an example that
we use to motivate and illustrate our framework.
In Section~\ref{sect:local-types}, we introduce a multiparty contract
model based on choreography synthesis.
In Section~\ref{sect:co2}, we present our version of \coco and
highlight its main features.
In Section~\ref{sec:honesty}, we define the notion of \emph{honesty}
and its practical importance in our contract-oriented scenario.
In Section~\ref{sec:result}, we present our results, which link the
notion of honesty to the progress and safety of a \coco system
(due to lack of space, the proofs are relegated to an online
appendix~\cite{OnlineAppendix}).
Finally, we discuss related work and conclude in Section~\ref{sec:conc}.

\subsection{A motivating example}
\label{sec:intro:example} \label{sec:example}
In this section, we introduce a running example to illustrate our framework.
We use $\ptpc{A}, \ptpc{B},\ldots$ for participant names, and $\ptpv{a},
\ptpv{a}', \ptpv{b}, \ldots$ for participant variables,
and use the colour \clc{blue} to highlight contracts.

Consider the following distributed scenario: an online
store $\ptpc{A}$ allows two buyers $\ptpv{b}_1$ and
$\ptpv{b}_2$ to make a joint purchase through a simplified protocol:
after they both $\atom{req}$uest the same item, a $\atom{quote}$ is
sent to $\ptpv{b}_1$, who is then expected to either place an
order ($\atom{order}$) or end the session ($\atom{bye}$); the store also
promises to notify $\ptpv{b}_2$ about whether the order was placed
($\atom{ok}$) or cancelled ($\atom{bye}$).
$\ptpc{A}$'s behaviour is described by the following contract:
\[
\clc{c_{\ptpc{A}}} \;=\; \clc{
  \PRECEIVE{b_1}{req} \PSEP \PRECEIVE{b_2}{req} \PSEP
  \PSEND{b_1}{quote} \PSEP
  \left(
    \PRECEIVE{b_1}{order} \PSEP \PSEND{b_2}{ok}
    \POUTCH
    \PRECEIVE{b_1}{bye} \PSEP \PSEND{b_2}{bye}
  \right)
}
\]
What kind of contracts would be compliant with
$\clc{c_{\ptpc{A}}}$?  One answer consists in the following contracts,
advertised by buyers $\ptpc{B}_1$ and $\ptpc{B}_2$. 
\begin{align*}
  \clc{c_{\ptpc{B}_1}} &\,=\; \clc{\PSEND{a}{req} \PSEP \PRECEIVE{a}{quote} \PSEP
    \left(
      \PSEND{b_2'}{ok} \PSEP \PSEND{a}{order} 
      \PINCH
      \PSEND{b_2'}{bye} \PSEP \PSEND{a}{bye}
    \right)
  } \\
  \clc{c_{\ptpc{B}_2}} &\,=\; \clc{\PSEND{a'}{req} \PSEP
    \left(
      \PRECEIVE{b_1'}{ok} \PSEP \PRECEIVE{a'}{ok}
      \POUTCH
      \PRECEIVE{b_1'}{bye} \PSEP \PRECEIVE{a'}{bye}
    \right)
  }
\end{align*}

Here,
$\ptpc{B}_1$ promises to send the $\atom{req}$uest to the store
($\ptpv{a}$), wait for the $\atom{quote}$, and then notify the other
buyer ($\ptpv{b_2'}$) before accepting or rejecting the store offer;
symmetrically, $\ptpc{B}_2$'s contract sends the $\atom{req}$uest to
the store ($\ptpv{a'}$), and then expects to receive the same
notification (either $\atom{ok}$ or $\atom{bye}$) from both the other
buyer ($\ptpv{b_1'}$) and the store itself.
%
Each contract represents the local viewpoint of the
participant who advertises it: $\clc{c_{\ptpc{A}}}$ represents
the local viewpoint of the store, and thus it does not (and indeed, it
cannot) capture the communications between $\ptpc{B}_1$ and
$\ptpc{B}_2$.

An agreement among $\clc{c_{\ptpc{A}}}$, $\clc{c_{\ptpc{B}_1}}$ and
$\clc{c_{\ptpc{B}_2}}$ may be found by replacing the participant
variables in each contract with actual names, e.g., with substitutions
$\subst{\ptpv{a},\ptpv{a}'}{\ptpc{A}}$, $\subst{\ptpv{b}_1,
  \ptpv{b}_1'}{\ptpc{B}_1}$ and $\subst{\ptpv{b}_2,
  \ptpv{b}_2'}{\ptpc{B}_2}$.  Such an agreement is based on the
existence of the following choreography (i.e., global type), which can be
synthesised similarly to what is done in~\cite{LT12}:
\[
\begin{array}{lcl}
  \GA_{\ptpc{A}\ptpc{B}_1\ptpc{B}_2} & = & 
  \GSEND{B_1}{A}{req} \GSEP
  \GSEND{B_2}{A}{req} \GSEP
  \GSEND{A}{B_1}{quote} \GSEP
  \\
  &&
  \left(
    \GSEND{B_1}{B_2}{ok} \GSEP 
    \GSEND{B_1}{A}{order} \GSEP
    \GSEND{A}{B_2}{ok}
    \;\;\; \GCH \;\;\;
    \GSEND{B_1}{B_2}{bye} \GSEP 
    \GSEND{B_1}{A}{bye} \GSEP 
    \GSEND{A}{B_2}{bye} 
  \right)
\end{array}
\]
The ability to synthesise $\GA_{\ptpc{A}\ptpc{B}_1\ptpc{B}_2}$
guarantees that the global type is well-formed and projectable
back to the initial contracts $\clc{c_{\ptpc{A}}}$, $\clc{c_{\ptpc{B}_1}}$ and
$\clc{c_{\ptpc{B}_2}}$ (with the substitutions above); this, in turn,
guarantees progress and safety~\cite{LT12} of the contractual agreement.
%

However, in a realistic scenario, the existence of a contractual
agreement among participants does not guarantee that progress and
safety will also hold at runtime: in fact, a participant may
advertise a contract promising some behaviour, and then fail to
respect it --- either maliciously or accidentally.
Such failure may then cascade on other participants, e.g., if they remain
stuck waiting for a promised message that is never sent.

This sort of situations can be modelled using the \coco calculus.
A \coco system for the store-and-two-customers example may be implemented as
follows:
\begin{align*}
  S_1 &\,\;=\;\; {(x,y,z)\big(
    {\cocoSys{\ptpc{A}}{\cocoTell{\ptpc{A}}{x}{c_{\ptpc{A}}}
        \cocoSeq \cocoFuse \cocoSeq P_{\ptpc{A}}}}
    \ \ \cocoPar\ \ 
    {\cocoSys{\ptpc{B}_{1}}{\cocoTell{\ptpc{A}}{y}{c_{\ptpc{B}_{1}}}
        \cocoSeq P_{\ptpc{B}_{1}}}}
    \ \ \cocoPar\ \ 
    {\cocoSys{\ptpc{B}_{2}}{\cocoTell{\ptpc{A}}{z}{c_{\ptpc{B}_{2}}}
        \cocoSeq P_{\ptpc{B}_{2}}}}
    \big)}
\end{align*}
Here, participant $\ptpc{A}$ advertises its contract
$\clc{\PV_\ptpc{A}}$ to itself via the primitive
$\cocoTell{\ptpc{A}\!}{x}{c_{\ptpc{A}}}$, where $x$ is used as a
session handle for interacting with other participants.  $\ptpc{B}_1$
and $\ptpc{B}_2$ advertise their respective contracts to $\ptpc{A}$
with a similar invocation.

In this example, $\ptpc{A}$ also plays the role of \emph{contract
  broker}: once all contracts have been advertised, the $\cocoFuse$
prefix can establish a new session, based on the fact that the global
agreement $\GA_{\ptpc{A}\ptpc{B}_1\ptpc{B}_2}$ can be synthesised from
$\clc{c_{\ptpc{A}}}$, $\clc{c_{\ptpc{B}_1}}$ and
$\clc{c_{\ptpc{B}_2}}$.
This new session is shared among participants
$\ptpc{A}$, $\ptpc{B}_1$ and $\ptpc{B}_2$.

At this point, the execution of the system (i.e., the continuation of
processes $P_{\ptpc{A}}$, $P_{\ptpc{B}_{1}}$ and $P_{\ptpc{B}_{2}}$)
is not required to respect the contracts.
In fact, we will see that when the contracts are violated, the
calculus allows for \emph{culpable} participants to be always ruled
out.
Furthermore, we will discuss \emph{honesty}, i.e., the guarantee
that a participant will always fulfil its advertised contracts ---
even in contexts where other participants fail to fulfil theirs.
When such a guarantee holds, the contractual progress and safety are
also reflected in the runtime behaviour of the \coco system.

\paragraph{Other possible agreements.}
Our contract model allows for other scenarios. For instance, a
participant $\ptpc{B}_{12}$ may impersonate both customers, and
promise to always accept the store offer, by advertising the following
contract:
\[
\clc{c_{\ptpc{B}_{12}}} \;=\; \PSEND{a''}{req} \PSEP \PSEND{a''}{req}
\PSEP \PRECEIVE{a''}{quote} \PSEP
\PSEND{a''}{order} \PSEP \PRECEIVE{a''}{ok}
\]
where the $\atom{req}$uest to the store ($\ptpv{a''}$) is sent twice
(i.e., once for each impersonated customer).
In this case, if we combine $\clc{c_{\ptpc{A}}}$ and
$\clc{c_{\ptpc{B}_{12}}}$ with substitutions
$\subst{\ptpv{a}''}{\ptpc{A}}$, $\subst{\ptpv{b}_1,
  \ptpv{b}_2}{\ptpc{B}_{12}}$, we can find an agreement by
synthesising the following global type:
\[
\begin{array}{lcl}
  \GA_{\ptpc{A}\ptpc{B}_{12}} & = & 
  \GSEND{B_{12}}{A}{req} \GSEP
  \GSEND{B_{12}}{A}{req} \GSEP
  \GSEND{A}{B_{12}}{quote} \GSEP
  \GSEND{B_{12}}{A}{order} \GSEP
  \GSEND{A}{B_{12}}{ok}
\end{array}
\]
Similarly to the previous case, this scenario may be modelled with the
following \coco system:
\begin{align*}
  S_2 &\,\;=\;\; {(x,w)\big(
    {\cocoSys{\ptpc{A}}{\cocoTell{\ptpc{A}}{x}{c_{\ptpc{A}}}
        \cocoSeq \cocoFuse \cocoSeq P_{\ptpc{A}}}}
    \ \ \cocoPar\ \ 
    {\cocoSys{\ptpc{B}_{12}}{\cocoTell{\ptpc{A}}{w}{c_{\ptpc{B}_{12}}}
        \cocoSeq P_{\ptpc{B}_{12}}}}
    \big)}
\end{align*}
where the $\cocoFuse$ prefix can now create a session involving
$\ptpc{A}$ and $\ptpc{B}_{12}$.

The participants in the \coco systems $S_1$ 
and $S_2$ 
may also be combined, so to obtain:
\begin{IEEEeqnarray*}{rCll}
  S_{12} &\,\;=\;\;& (x,y,z,w)\big(\;&
  {\cocoSys{\ptpc{A}}{\cocoTell{\ptpc{A}}{x}{c_{\ptpc{A}}}
      \cocoSeq \cocoFuse \cocoSeq P_{\ptpc{A}}}} \ \cocoPar\ \ 
  {\cocoSys{\ptpc{B}_{1}}{\cocoTell{\ptpc{A}}{y}{c_{\ptpc{B}_{1}}}
      \cocoSeq P_{\ptpc{B}_{1}}}}
  \ \ \cocoPar\ \ 
  {\cocoSys{\ptpc{B}_{2}}{\cocoTell{\ptpc{A}}{z}{c_{\ptpc{B}_{2}}}
      \cocoSeq P_{\ptpc{B}_{2}}}} \\
  & & & \ \cocoPar\ \ 
  {\cocoSys{\ptpc{B}_{12}}{\cocoTell{\ptpc{A}}{w}{c_{\ptpc{B}_{12}}}
      \cocoSeq P_{\ptpc{B}_{12}}}}
  \;\big)
\end{IEEEeqnarray*}
In this case, after all contracts have been advertised to $\ptpc{A}$,
either a session corresponding to
$\GA_{\ptpc{A}\ptpc{B}_1\ptpc{B}_2}$, or to
$\GA_{\ptpc{A}\ptpc{B}_{12}}$ may take place, thus involving a
different number of participants depending on which contracts are
$\cocoFuse$d.
In such cases, it makes sense to consider whether one of the agreements
should take precedence over the other, and which criteria should drive
this choice.

%% file: session.tex
\section{A Choreography-Based Contract Model}
\label{sect:local-types}
We introduce a contract model based on concepts and
results from the session types literature.
Individual contracts are expressed using the syntax of local session types;
while contractual compliance is based on global types synthesis:
a set of contracts is compliant if it is possible to
synthesise a choreography from it, as described in \cite{LT12}.
For simplicity, we adopt
syntax and semantics in the style of \cite{cdp11,DY12}: we use
participant names (instead of channels) for message exchange, i.e.,
we consider systems with just one channel between each pair of
participants.

\paragraph{Syntax \& Semantics.}
Let $\ptpNames$ and $\ptpVars$ be disjoint sets of, respectively,
\emph{participant names} (ranged over by $\ptpc{A}, \ptpc{B}, \ldots$)
and \emph{participant variables} (ranged over by $\ptpv{a}, \ptpv{b},
\ldots$).  Let $\ptpcv{a}, \ptpcv{b}$ range over $\ptpNames \cup
\ptpVars$.
The syntax of contracts below is parametrised wrt \emph{sorts}
(ranged over by $\SORTV$) which abstract data types (either simple or complex).
We use the colour \clc{blue} for single contracts and \clsy{green} for
\emph{systems} of contracts.
\[
\begin{array}
  {l@{\quad}l@{\quad}c@{\hspace{1em}}c@{\hspace{1em}}c@{\hspace{1em}}c@{\hspace{1em}}c@{\hspace{1em}}c@{\hspace{1em}}c}
  \SYSV, \clsy{\SYSV'} & ::= & \clsy{\SYSV \PPAR \SYSV'}
  & \BNFOR & \PBOX{A}{\PV}
  & \BNFOR & \QUEUEN{\rho}
  & \BNFOR & \END
  \\[0.1pc]
  \PV, \PV'  & ::= & 
  \PINS{i \in I}{\PSEND{\ptpcv{a}_i}{\SORTV_i}{\PSEP \PV_i}}
  & \BNFOR &
  \PSUM{i \in I}{\PRECEIVE{\ptpcv{a}}{\SORTV_i} \PSEP \PV_i}
  & \BNFOR & \PRECN \PV
  & \BNFOR & \PRECVN 
\end{array}
\]
A contract $\PV$ may be either:
($i$) an internal choice $\clc{\PINCH}$,
with the intuitive semantics that after sending the message $\SORTV_i$
to participant $\ptpcv{a}_i$, behaviour $\clc{\PV_i}$ take places;
($ii$) an external choice $\clc{\sum}$, saying that if a message of sort
$\SORTV_i$ is received from $\ptpcv{a}$, then
behaviour $\clc{\PV_i}$ takes place;
or ($iii$) a recursive behaviour.  We
assume that $\forall i \neq j \in I \suchthat (\ptpcv{a}_i,\SORTV_i)
\neq (\ptpcv{a}_j,\SORTV_j)$ in internal choices,
$\forall i \neq j \in I \suchthat \SORTV_i
\neq \SORTV_j$ in
external choices, and
that $\clc{\PINCH}$ and $\clc{\POUTCH}$ are associative and commutative.
We write $\fv{\PV}$ for the free participant variables in $\PV$.

A system of contracts $\SYSV$ may be either: ($i$) a parallel
composition of systems $\SYSV \PPAR \clsy{\SYSV'}$; or
($ii$) a \emph{named} contract $\PBOX{A}{\PV}$, saying that participant
$\ptpc{A}$ promises to behave according to $\clc{c}$;
($iii$) a \emph{queue} $\QUEUEN{\rho}$ of messages from $\ptpc{A}$ to
$\ptpc{B}$.
In a system $\SYSV$, we assume that there is at most one queue per pair of
participants, (i.e., one channel per direction), and that participant
names are pairwise distinct.

\label{par:session:semantics}
We consider systems of contracts as processes whose semantics is
given by the following main reduction rules (see the online appendix \cite{OnlineAppendix} for the omitted ones):
\[
\begin{array}{rcl}
  \PBOX{A}{\PSEND{B}{\SORTV} \PSEP \PV_0 \PINCH \PV_1}
  \PPAR \QUEUE{AB}{\rho}
  \PPAR \SYSV
  & \semSend{\ptpc A}{e}{\ptp B} &
  \PBOX{A}{\PV_0}
  \PPAR \QUEUE{AB}{\rho \cdot \SORTV}
  \PPAR \SYSV
  \\
  \PBOX{A}{\PRECEIVE{B}{\SORTV} \PSEP \PV_0 \POUTCH \PV_1}
  \PPAR \QUEUE{BA}{\SORTV \cdot \rho}
  \PPAR \SYSV
  & \semRecv{\ptpc A}{e}{\ptpc B} &
  \PBOX{A}{\PV_0}
  \PPAR \QUEUE{BA}{\rho}
  \PPAR \SYSV
\end{array}
\]
The first rule says that, after an internal choice, participant $\ptpc A$
puts a message $\SORTV$ on its queue for participant $\ptpc B$.
The second rule says that $\ptpc{A}$'s external choice can receive a
message of the right sort from an input queue $\ptpc{BA}$.
%
We write $\SYSV
\cocoCmove{A}{e}{B} \clsy{\SYSV'}$ when either $\SYSV
\semSend{\ptpc{A}}{e}{\ptpc{B}} \clsy{\SYSV'}$ or $\SYSV
\semRecv{\ptpc{A}}{e}{\ptpc{B}} \clsy{\SYSV'}$, and
$\QUEUES{\SYSV}$ for the parallel composition of the empty
queues connecting all pairs of participants in $\SYSV$.

\begin{example} \label{ex:session:reductions}
  From the example in Section~\ref{sec:intro:example}, consider the
  instantiated contracts of the store $\ptpc{A}$ and its customer
  $\ptpc{B}_{12}$.  We illustrate the initial system, and how it
  progresses:
  \[
  \setlength\arraycolsep{1pt}
  \begin{array}{lclllllll}
    \clsy{\SYSV_{\ptpc{A}\ptpc{B}_{12}Q}} &=&
    \PBOX{\ptpc{A}}{c_{\ptpc{A}}\subst{\ptpv{a''}}{\ptpc{A}}
      \subst{\ptpv{b}_1,\ptpv{b}_2}{\ptpc{B}_{12}}} 
    &\PPAR&
    \PBOX{\ptpc{B}_{12}}{c_{\ptpc{B}_{12}}\subst{\ptpv{a''}}{\ptpc{A}}
      \subst{\ptpv{b}_1,\ptpv{b}_2}{\ptpc{B}_{12}}}
    &\PPAR& \QUEUE{AB_{12}}{\emptyQ}
    &\PPAR &\QUEUE{B_{12}A}{\emptyQ}
    \\[0.2em]
    &=&
    \PBOX{\ptpc{A}}{
      \clc{
        \PRECEIVE{B_{12}}{req} \PSEP \PRECEIVE{B_{12}}{req} \PSEP
        \ldots
      }
    } &\PPAR&
    \PBOX{\ptpc{B}_{12}}{
      \PSEND{A}{req} \PSEP \PSEND{A}{req}
      \PSEP
      \ldots
    }
    &\PPAR &\QUEUE{AB_{12}}{\emptyQ}
    & \PPAR &\QUEUE{B_{12}A}{\emptyQ}
    \\[0.2em]
    \multicolumn{2}{l}{\semSend{\ptpc{B_{12}}}{req}{\ptp{A}}}
    &
    \PBOX{\ptpc{A}}{
      \clc{
        \PRECEIVE{B_{12}}{req} \PSEP
        \PRECEIVE{B_{12}}{req} \PSEP
        \ldots
      }
    } &\PPAR&
    \PBOX{\ptpc{B}_{12}}{
      \PSEND{A}{req}
      \PSEP
      \ldots
    }
    &\PPAR &\QUEUE{AB_{12}}{\emptyQ}
    & \PPAR &\QUEUE{B_{12}A}{\msgsort{req}}
    \\[0.2em]
    \multicolumn{2}{l}{ \semRecv{\ptpc{A}}{req}{\ptpc {B_{12}}} \qquad \qquad}
    &
    \PBOX{\ptpc{A}}{
      \clc{
        \PRECEIVE{B_{12}}{req} \PSEP
        \ldots
      }
    } &\PPAR&
    \PBOX{\ptpc{B}_{12}}{
      \PSEND{A}{req}
      \PSEP
      \ldots
    }
    &\PPAR &\QUEUE{AB_{12}}{\emptyQ}
    & \PPAR &\QUEUE{B_{12}A}{\emptyQ}
  \end{array}
  \]
\end{example}

\paragraph{Choreography Synthesis as Compliance.}
We briefly introduce the compliance relation that tells
whether some contracts can be combined to describe a correct interaction.
We reuse the main results from~\cite{LT12}: a typing system which assigns
a (unique) global type to a set of local types.
We say that a set of contracts (i.e., local types) is \emph{compliant} if it can be
assigned a choreography, i.e., a global type.

For simplicity, we use only a subset of the global
types originally supported (we conjecture that extending this
would not pose any difficulties).
The main difference
is that, in the style of \cite{cdp11,DY12}, we replace channels with
participant names.
%

The syntax of global types is as follows:
\[\begin{array}{c}
  \GA \ \ \ \bnfdef \ \ \ \GSEND{A}{B}{\SORTV} \GSEP \GA
  \BNFOR \GA \GCH \GA'
  \BNFOR \GA \GPAR \GA'
  \BNFOR  \GRECN \GA
  \BNFOR \GRECV
  \BNFOR \GEND
\end{array}\]
where the first production means that a participant $\ptpc A$ sends
a message of sort $\SORTV$ to $\ptpc B$, then interactions in $\GA$ take place;
$\GA \GCH \GA'$ means that either interactions in $\GA$, or
in $\GA'$ take place; $\GA \GPAR \GA'$ means that interactions in
$\GA$ and $\GA'$ are executed concurrently; the rest of the
productions are for recursive interactions, and end.

Similarly to the original synthesis, we use judgements of the form
$\judge{\Gamma}{\SYSV}{\GA}$, where $\Gamma$ is an environment to keep track
of recursion variables, $\SYSV$ is a system of contracts, and $\GA$ is the 
global type assigned to $\SYSV$.
We say that a system of contracts \emph{$\SYSV$ has global type $\GA$}, if
one can infer the judgement 
$\judge{\emptyctx}{\SYSV}{\GA}$ from the rules in the online appendix \cite{OnlineAppendix}
(simplified from \cite{LT12}) where $\emptyctx$ is the empty context $\Gamma.$ 
Essentially, the synthesis rules allow to execute a set of contracts step-by-step,
while keeping track of the structure of the interactions in a global type.

The main properties that we are interested in
--- and that are guaranteed
by the synthesis ---
is that the inferred global type is $(i)$ well-formed, and $(ii)$ projectable back
to the original  contracts.
Essentially, this means that each local type must be single-threaded,
and that \emph{knowledge of choice} is preserved --- i.e., each choice
is made by exactly one participant, and all the others are either made
aware of the choice, or they have the same behaviour whatever choice
is made.
%

\def\caSubstReductChoice{
  \PRECEIVE{B_1}{order} \PSEP \PSEND{B_2}{ok}
  \POUTCH
  \PRECEIVE{B_1}{bye} \PSEP \PSEND{B_2}{bye}
}
\def\caSubstReduct{
  \PRECEIVE{B_2}{req} \PSEP
  \PSEND{B_1}{quote} \PSEP
  \left(
    \caSubstReductChoice
  \right)
}
\def\caSubst{
  \PRECEIVE{B_1}{req} \PSEP \caSubstReduct
}
\def\cboneSubstReductChoice{
    \PSEND{B_2}{ok} \PSEP \PSEND{A}{order}
    \PINCH
    \PSEND{B_2}{bye} \PSEP \PSEND{A}{bye}
}
\def\cboneSubstReduct{
  \PRECEIVE{A}{quote} \PSEP
  \left(
    \cboneSubstReductChoice
  \right)
}
\def\cboneSubst{
  \PSEND{A}{req} \PSEP \cboneSubstReduct
}
\def\cbtwoSubstReductChoice{
  \PRECEIVE{B_1}{ok} \PSEP \PRECEIVE{A}{ok}
  \POUTCH
  \PRECEIVE{B_1}{bye} \PSEP \PRECEIVE{A}{bye}
}
\def\cbtwoSubst{
\PSEND{A}{req} \PSEP
      \left(
        \cbtwoSubstReductChoice
      \right)
}
\begin{example} \label{ex:session:judgement}
  Building up on the example from Section~\ref{sec:intro:example}, we
  combine the contract of store $\ptpc{A}$ with those of customers
  $\ptpc{B}_1$ and $\ptpc{B}_2$, and we obtain the system:
  \[
  \begin{array}{rcl}
    \clsy{\SYSV_{\ptpc{A}\ptpc{B}_1\ptpc{B}_2}} &\;=\;&
    \PBOX{\ptpc{A}}{c_{\ptpc{A}}\subst{\ptpv{b}_1}{\ptpc{B}_1}
      \subst{\ptpv{b}_2}{\ptpc{B}_2}} \;\PPAR\;
    \PBOX{\ptpc{B}_1}{c_{\ptpc{B}_1}\subst{\ptpv{a}}{\ptpc{A}}
      \subst{\ptpv{b}'_2}{\ptpc{B}_2}} \;\PPAR\;
    \PBOX{\ptpc{B}_2}{c_{\ptpc{B}_2}\subst{\ptpv{a}'}{\ptpc{A}}
      \subst{\ptpv{b}'_1}{\ptpc{B}_1}}
    \\[.2pc]
    &\;=\;&
    \PBOX{\ptpc{A}}{
      \caSubst
    } \\[.2pc]
    & & \;\PPAR\;
    \PBOX{\ptpc{B}_1}{
      \cboneSubst
    } \\[.2pc]
    & & \;\PPAR\;
    \PBOX{\ptpc{B}_2}{
      \cbtwoSubst
    }
  \end{array}
  \]
  which can be assigned the following global type:
  \[
  \begin{array}{lcl}
    \GA_{\ptpc{A}\ptpc{B}_1\ptpc{B}_2} & = & 
    \GSEND{B_1}{A}{req} \GSEP
    \GSEND{B_2}{A}{req} \GSEP
    \GSEND{A}{B_1}{quote} \GSEP
    \\
    &&
    \left(
      \GSEND{B_1}{B_2}{ok} \GSEP 
      \GSEND{B_1}{A}{order} \GSEP
      \GSEND{A}{B_2}{ok}
      \;\;\; \GCH \;\;\;
      \GSEND{B_1}{B_2}{bye} \GSEP 
      \GSEND{B_1}{A}{bye} \GSEP 
      \GSEND{A}{B_2}{bye} 
    \right)
  \end{array}
  \]
  that is to say that  $\judge{\emptyctx}{\clsy{\SYSV_{\ptpc{A}\ptpc{B}_1\ptpc{B}_2}}}{\GA_{\ptpc{A}\ptpc{B}_1\ptpc{B}_2}}$ holds.
  Instead, if we combine the store $\ptpc{A}$ with $\ptpc{B}_{12}$ we have
  \[
  \begin{array}{rcl}
    \clsy{\SYSV_{\ptpc{A}\ptpc{B}_{12}}} &\;=\;&
    \PBOX{\ptpc{A}}{c_{\ptpc{A}}\subst{\ptpv{b}_1,\ptpv{b}_2}{\ptpc{B}_{12}}}
    \;\PPAR\;
    \PBOX{\ptpc{B}_{12}}{c_{\ptpc{B}_{12}}\subst{\ptpv{a}''}{\ptpc{A}}} \\[.2pc]
    &\;=\;&
    \PBOX{\ptpc{A}}{
      \PRECEIVE{B_{12}}{req} \PSEP \PRECEIVE{B_{12}}{req} \PSEP
      \PSEND{B_{12}}{quote} \PSEP
      \left(
        \PRECEIVE{B_{12}}{order} \PSEP \PSEND{B_{12}}{ok}
        \POUTCH
        \PRECEIVE{B_{12}}{bye} \PSEP \PSEND{B_{12}}{bye}
      \right)
    } \\[.2pc]
    & & \;\PPAR\;
    \PBOX{\ptpc{B}_{12}}{
      \PSEND{A}{req} \PSEP \PSEND{A}{req}
      \PSEP \PRECEIVE{A}{quote} \PSEP
      \PSEND{A}{order} \PSEP \PRECEIVE{A}{ok}
    }
    \\[0.75em]
    \GA_{\ptpc{A}\ptpc{B}_{12}} & = & 
    \GSEND{B_{12}}{A}{req} \GSEP
    \GSEND{B_{12}}{A}{req} \GSEP
    \GSEND{A}{B_{12}}{quote} \GSEP
    \GSEND{B_{12}}{A}{order} \GSEP
    \GSEND{A}{B_{12}}{ok}
  \end{array}
  \]
  and, again, the judgement
  $\judge{\emptyctx}{\clsy{\SYSV_{\ptpc{A}\ptpc{B}_{12}}}}{\GA_{\ptpc{A}\ptpc{B}_{12}}}$
  holds.
\end{example}


%% file: co2.tex
\section{A Multiparty Version of \coco} \label{sect:co2}

We introduce a version of the \coco calculus (for COntract-Oriented
computing)~\cite{BTZ12} adapted to multiparty contracts and sessions.
Let $\sesNames$ and $\sesVars$ be disjoint sets of, respectively,
\emph{session names} (ranged over by $\sesc{S}, \sesc{S'}, \ldots$)
and \emph{session variables} (ranged over by $\sesv{x}, \sesv{y},
\sesv{z} \ldots$).  Let $\sesv{u}, \sesv{v}, \ldots$ range over
$\sesNames \cup \sesVars$.

\paragraph{Syntax \& Semantics.}
The syntax of \coco is given by the following productions:
\[\begin{array}{lrcl}
  \text{Processes} & P,Q  & \bnfdef & \cocoSum_{i \in I} p_i \cocoSeq P_i
  \bnfsep    P \cocoPar Q
  \bnfsep    (\vec{u}, \vec{\ptpv{a}})P
  \bnfsep    X(\vec u,\vec{\ptpcv{a}})
  \bnfsep    \cocoEmptyproc
  \\[.2pc] 
  \text{Prefixes} & p
  & \bnfdef & \cocoTau
  \bnfsep    \cocoTell{\,\ptpcv{a}\!}{u}{\PV}
  \bnfsep    \cocoFuse
  \bnfsep    \cocoDo{u}{\ptpcv a}{e}
  \\[.2pc] 
  \text{Latent contracts} & K  & \bnfdef &
  {\cocoFreeze u {\pmv A \cocoSays \PV}} \bnfsep K \cocoPar K
  \\[.2pc]
  \text{Systems} & S  & \bnfdef &
  \cocoSys {\pmv A} P \bnfsep 
  \cocoSys {\pmv A} K \bnfsep 
  \cocoSys {\sesc s} \SYSV \bnfsep S \cocoPar S \bnfsep
  (\vec{\sesv{u}}, \vec{\ptpv{a}})S
  \bnfsep \cocoEmptysys
\end{array}\]
\coco features CCS-style processes, equipped with branching $\cocoSum$
(not to be confused with the choice operator used in contracts), parallel
composition $\cocoPar$, restrictions of session and participant variables,
and named process invocation.
The prefixes are for internal action ($\cocoTau$),
contract advertisement ($\cocoTell{\!\!\!}{}{\!\!\!}$), session
creation upon contractual agreement ($\cocoFuse$), and execution of
contractual actions ($\cocoDo{}{}{\!}$).
A latent contract of the form $\cocoFreeze u {\pmv A \cocoSays \PV}$
represents the promise of participant $\ptpc{A}$ to fulfil $\PV$ by
executing $\cocoDo{}{}{\!}$-actions on a session variable $u$.
\coco systems may be parallel compositions of processes $\cocoSys {\pmv A}
P$ (where $\ptpc A$ is the participant executing $P$), \emph{latent}
contracts $\cocoSys {\pmv A} K$ (where $\ptpc{A}$ is the participant
to which the contracts in $K$ have been advertised), and established
sessions $\cocoSys {\sesc s} \SYSV$ (where $s$ is a session name, and
$\SYSV$ is a system of \emph{stipulated} contracts as in Section~\ref{sect:local-types}). 
We assume well-formed systems where each participant $\pmv{A}$ has at
most one process $\cocoSys{\pmv{A}}{P}$.
Note that \coco process and system
productions allow to delimit both session names/variables ($\vec u$)
and participant variables ($\vec{\ptpv{a}}$), but \emph{not}
participant names, which are considered public.

\label{par:co2:semantics}

We give the main reduction rules for the semantics of \coco
(where $\cocoPlus$ and $\cocoPar$ are standard associative and commutative operators):

\[
\begin{array}{c}
  \ltsruleS{\coco-Tell}
  {}
  {\cocoSys {\pmv A} {\cocoTell{B}{x}{\PV}
      \cocoSeq P \cocoPlus P' \cocoPar Q}
    \ \cocoMove{}\ 
    \cocoSys {\pmv A} {P \cocoPar Q} \ \cocoPar\ 
    \cocoSys {\pmv B} {\cocoFreeze x {\pmv A \cocoSays \PV}}
  }
\end{array}
\]
\[
\begin{array}{c}
  \ltsrule{\coco-Fuse}
  {\cocoAgreement{\sigma}{\pi}{K}{\SYSV}
    \\ \vec {\ptpv a} = \dom \pi
    \\ \vec u = \dom \sigma
    \\ \img{\sigma} = \setenum{s}
    \\ s\text{ fresh}
  }
  {(\vec u, \vec {\ptpv a})\left( \cocoSys {\pmv A}
      {\cocoFuse \cocoSeq P \cocoPlus P' \cocoPar Q}
      \ \cocoPar\ \cocoSys {\pmv A} {K}
      \ \cocoPar\ S \right)
    \ \cocoMove{}\ 
    (s)\left( \cocoSys {\pmv A} {P \cocoPar Q}\sigma\pi
      \ \cocoPar\ \cocoSys s {
        \SYSV \PPAR \QUEUES{\SYSV}
      } \ \cocoPar\ S\sigma\pi \right)
  }
\end{array}
\]
\[
\begin{array}{c}
  \ltsrule{\coco-Do}
  {\SYSV \cocoCmove{A}{e}{B} \clsy{\SYSV'}}
  {\cocoSys s {\SYSV} \ \cocoPar \ 
    \cocoSys {\ptpc A} {\cocoDo s {\ptpc B} {e} \cocoSeq P
      \cocoPlus P' \cocoPar Q}
    \ \cocoMove{}\ 
    \cocoSys s {\clsy{\SYSV'}} \ \cocoPar \ 
    \cocoSys {\ptpc A} {P \cocoPar Q}
  }
\end{array}
\]
$\rulename{\coco-Tell}$ allows a participant $\ptpc{A}$ to advertise a
contract $\PV$ to $\ptpc{B}$; as a result, a new \emph{latent} contract is
created, recording the fact that it was promised by $\ptpc{A}$.
$\rulename{\coco-Fuse}$ establishes a new session: the latent
contracts held in $\cocoSys {\ptpc A} {K}$ are combined, and their
participant variables substituted, in order to find an
\emph{agreement}, i.e., a $\SYSV$ which satisfies the relation
$\cocoAgreement{\sigma}{\pi}{K}{\SYSV}$ (see
Definition~\ref{def:co2:agreement} below).
Provided an agreement is found, fresh session variable $s$ and
participants names are
shared among the parties, via substitutions $\sigma$ and $\pi$;
within the session, the involved contracts become \emph{stipulated}
(as opposed to ``latent'', before the agreement).
Rule $\rulename{\coco-Do}$ allows $\ptpc{A}$ to perform an
input/output action $\atom{e}$ towards $\ptpc{B}$ on session $s$,
provided that $\SYSV$ permits it.
The omitted \coco rules are standard, and they are listed in
the online appendix \cite{OnlineAppendix}:
they cover internal actions, parametric
processes, parallel composition, and delimitations.

When needed, we label \coco system transitions: $S \xrightarrow{{\ptpc
    A} \colon p} S'$ means that $S$ reduces to $S'$ through a prefix
$p$ fired by participant $\ptpc{A}$.

\begin{example} \label{ex:co2:do}
  Consider the \coco system:
  \[
  S \;=\; {\cocoSys{\ptpc{A}}{
      \cocoDo s {\ptpc B} {int}
      \cocoPlus
      \cocoDo s {\ptpc B} {bool}   
    }
  }
  \ \ \cocoPar\ \ 
  \cocoSys s {
    \PBOX{A}{\PSEND{B}{int}}
    \PPAR
    \PBOX{B}{\PRECEIVE{A}{int}}
    \PPAR
    \QUEUE{AB}{\emptyQ}
    \PPAR
    \QUEUE{BA}{\emptyQ}
  }
  \ \ \cocoPar\ \ 
  {\cocoSys{\ptpc{B}}{
      \cocoDo s {\ptpc A} {int}
    }
  }
  \]
  Here, the \coco process of participant $\ptpc{A}$ can perform an
  action towards $\ptpc{B}$ on session $s$, with either a message of
  sort $\atom{int}$ or $\atom{bool}$.
  However, $\ptpc{A}$'s contract in $s$ only specifies that
  $\ptpc{A}$ should send a message of sort $\atom{int}$ to $\ptpc{B}$:
  therefore, according to rule $\rulename{\coco-Do}$, only the first
  branch of $\ptpc{A}$ may be chosen, and the system reduces as follows.
  \[
  \begin{array}{rcccccc}
    S &\ \ \ \ \xrightarrow{{\ptpc A} \colon \cocoDo{s}{\ptpc B}{int}}\ \ \ \ &
    {\cocoSys{\ptpc{A}}{
        \cocoEmptyproc
      }
    }
    &\ \cocoPar\ &
    \cocoSys s {
      \PBOX{A}{\PEND}
      \PPAR
      \PBOX{B}{\PRECEIVE{A}{int}}
      \PPAR
      \QUEUE{AB}{\msgsort{int}}
      \PPAR
      \QUEUE{BA}{\emptyQ}
    }
    &\ \cocoPar\ &
    {\cocoSys{\ptpc{B}}{
        \cocoDo s {\ptpc A} {int}
      }
    } \\
    &\ \ \ \ \xrightarrow{{\ptpc B} \colon \cocoDo{s}{\ptpc A}{int}}\ \ \ \ &
    {\cocoSys{\ptpc{A}}{
        \cocoEmptyproc
      }
    }
    &\ \cocoPar\ &
    \cocoSys s {
      \PBOX{A}{\PEND}
      \PPAR
      \PBOX{B}{\PEND}
      \PPAR
      \QUEUE{AB}{\emptyQ}
      \PPAR
      \QUEUE{BA}{\emptyQ}
    }
    &\ \cocoPar\ &
    {\cocoSys{\ptpc{B}}{
        \cocoEmptyproc
      }
    }
  \end{array}
  \]
\end{example}

A main difference between our adaptation of \coco and the original
presentation comes from the way we specify session establishment.
We adopt the session agreement relation defined below.
\begin{definition}[Agreement relation $\cocoAgreement{\sigma}{\pi}{K}{\SYSV}$]
  \label{def:co2:agreement}
  Let
  \(
  K \equiv \PPROD{i \in I}{\cocoFreeze{x_i}{{\pmv A}_i \cocoSays \clc{\PV_i}}}
  \), such that
  \(
  \forall \, i \neq j \in I : \ptpc{A}_i \neq \ptpc{A}_j
  \),
  and let
  $\pi \colon \ptpVars \to \ptpNames$ and $\sigma \colon
  \sesVars \to \sesNames$ be two substitutions mapping
  participant variables to names, and session variables to names, respectively.
  Also, let
  \(
  \SYSV \equiv \clsy{\PPROD{i \in I}{\PBOX{A_i}{\PV_i}\pi}}
  \).
  We define:
  \[
  \cocoAgreement{\sigma}{\pi}{K}{\SYSV} 
  \iff
  \begin{array}{l}
    \dom{\sigma} = \bigcup_{i \in I} \setenum{x_i}
    \quad \land \quad
    \dom{\pi} = \bigcup_{i \in I} \fv{\clc{c_i}}
    \\
    \land
    \\
    \forall i \in I \suchthat \forall \ptpv{a} \in
    \fv{\clc{\PV_i}} \suchthat \pi(\ptpv{a}) \neq \ptpc{A}_i
    \quad \land \quad
    \exists \, \GA \suchthat \; \judge{\emptyctx}{ \SYSV }{\GA}
  \end{array}
  \]
\end{definition}
Intuitively, a system of stipulated contracts $\SYSV$ is constructed from a set
of latent contracts $K$, using a substitution $\pi$ that maps all the
participant variables in $K$ to the participant names in $K$ itself.
If it is possible to synthesise a
global type $\GA$ out of $\SYSV$, then the relation holds, and a
contractual agreement exists.
The first two conditions, on $\sigma$ and $\pi$, guarantee that all
the session and participant variables are indeed instantiated.
The third condition ensures that within a contract $\clc{\PV_i}$, belonging to
$\ptpc{A}_i$, no free participant variable in $\clc{\PV_i}$ is substituted by $\ptpc{A}_i$
itself.
Note that due to the condition imposed on $K$, each
participant may have at most one contract per session.

\def\piSubst{
  \setenum{\nicefrac{\ptpc{A}}{\ptpv{a},\ptpv{a}'},
    \nicefrac{\ptpc{B}_1}{\ptpv{b}_1, \ptpv{b}_1'},
    \nicefrac{\ptpc{B}_2}{\ptpv{b}_2,\ptpv{b}_2'}}
}
\begin{example} \label{ex:co2:agreement}
  We now illustrate how Definition~\ref{def:co2:agreement} works.
  Consider the following \coco system, with $\ptpc{A}$, $\ptpc{B}_{1}$,
  $\ptpc{B}_{2}$ from Section~\ref{sec:intro:example}, and
  $\clsy{\SYSV_{\ptpc{A}\ptpc{B}_1\ptpc{B}_2}}$ from
  Example~\ref{ex:session:judgement}:
  \[
  \!\!\!\!\!\begin{array}{rl}
    S_1 \;\;=& {(x,y,z)\big(
        {\cocoSys{\ptpc{A}}{\cocoTell{\ptpc{A}}{x}{c_{\ptpc{A}}}
            \cocoSeq \cocoFuse \cocoSeq P_{\ptpc{A}}}}
        \ \ \cocoPar\ \ 
        {\cocoSys{\ptpc{B}_{1}}{\cocoTell{\ptpc{A}}{y}{c_{\ptpc{B}_{1}}}
            \cocoSeq P_{\ptpc{B}_{1}}}}
        \ \ \cocoPar\ \ 
        {\cocoSys{\ptpc{B}_{2}}{\cocoTell{\ptpc{A}}{z}{c_{\ptpc{B}_{2}}}
            \cocoSeq P_{\ptpc{B}_{2}}}}
      \big)} \\[.4em]
    \cocoMove{}\cocoMove{}\cocoMove{}
    & {(x,y,z)\big(
        {\cocoSys{\ptpc{A}}{\cocoFuse \cocoSeq P_{\ptpc{A}}}}
        \ \cocoPar\ 
        {\cocoSys{\ptpc{A}}{\cocoFreeze x {\ptpc{A} \cocoSays \clc{\PV_{\ptpc{A}}}}
            \cocoPar\; \cocoFreeze y {\ptpc{B}_{1} \cocoSays \clc{\PV_{\ptpc{B}_{1}}}}
            \cocoPar\; \cocoFreeze z {\ptpc{B}_{2} \cocoSays \clc{\PV_{\ptpc{B}_{2}}}}}}
        \ \cocoPar\ 
        {\cocoSys{\ptpc{B}_{1}}{P_{\ptpc{B}_{1}}}}
        \ \cocoPar\ 
        {\cocoSys{\ptpc{B}_{2}}{P_{\ptpc{B}_{2}}}}
      \big)} \\[.4em]
    \cocoMove{\ptpc{A}\colon \cocoFuse} (s)S_1' =
    & {(s)\big(
        {\cocoSys{\ptpc{A}}{P_{\ptpc{A}}}}\sigma\pi
        \ \ \cocoPar\ \ 
        {\cocoSys{s}{
            \clsy{
              \SYSV_{\ptpc{A}\ptpc{B}_1\ptpc{B}_2}
              \cocoPar \QUEUES{\SYSV_{\ptpc{A}\ptpc{B}_1\ptpc{B}_2}}}
          }
        }
        \ \ \cocoPar\ \ 
        {\cocoSys{\ptpc{B}_{1}}{P_{\ptpc{B}_{1}}}}\sigma\pi
        \ \ \cocoPar\ \ 
        {\cocoSys{\ptpc{B}_{2}}{P_{\ptpc{B}_{2}}}}\sigma\pi
      \big)}
    \\[.6em]
    \multicolumn{2}{r}{
      \text{where } 
      \sigma = \subst{x,y,z}{s}
      \text{ and }
      \pi = \piSubst
    }
  \end{array}
  \]

  The initial system $S_1$ is the one considered in Section~\ref{sec:intro:example},
  where all the participants are ready to advertise their respective
  contracts to the store $\ptpc{A}$, by using a
  $\cocoTell{A\!\!\!}{\!\!}{\!\!}$-primitive.
  This has the effect of creating corresponding
  latent contracts within $\ptpc{A}$.
  Once all the latent contracts are in a same location, they may be
  $\cocoFuse$d.
  In this case, given $\sigma$ and $\pi$ as above,
  Definition~\ref{def:co2:agreement} is indeed applicable: the domains
  of $\sigma$ and $\pi$ comply with the definition's premises, and we already saw
  that a system consisting of $\clc{\PV_{\ptpc{A}}}$,
  $\clc{\PV_{\ptpc{B}_{1}}}$ and $\clc{\PV_{\ptpc{B}_{2}}}$ may be
  assigned a global type.
  Hence, a new session $s$ is created, based on the system of
  contracts $\clsy{\SYSV_{\ptpc{A}\ptpc{B}_1\ptpc{B}_2}}$, plus the queues
  connecting all pairs of participants.
  The session variables of the latent contracts being $\cocoFuse$d (i.e., $x$
  for participant $\ptpc{A}$, $y$ for $\ptpc{B}_1$, and $z$ for
  $\ptpc{B}_2$) are all substituted with the fresh session name $s$ in
  the processes $P_{\ptpc{A}}$, $P_{\ptpc{B}_1}$ and $P_{\ptpc{B}_2}$,
  via $\sigma$.
  Similarly for participant variables which are substituted
  with participant names, via $\pi$.
\end{example}

The \coco semantic rules are to be considered up-to a standard
structural congruence relation $\equiv$ (cf.\
\cite{OnlineAppendix}): we just point out that
$\cocoSys{\pmv A}{K} \cocoPar \cocoSys{\pmv A}{K'} \equiv \cocoSys
{\pmv A} {K \cocoPar K'}$ allows to select a compliant subset
from a group of latent contracts, before performing a $\cocoFuse$ ---
thus adding flexibility to the synthesis of choreographies.

\begin{example}
  Consider the system:
  \[
  \ldots
  \cocoSys {\ptpc{B}} {\cocoFuse \cocoSeq P \cocoPar Q}
  \ \ \cocoPar\ \ 
  \cocoSys {\ptpc{B}} {
    \cocoFreeze x {\pmv A_1 \cocoSays \PSEND{a}{int}}
    \ \cocoPar\ 
    \cocoFreeze y {\pmv A_2 \cocoSays \PRECEIVE{a'}{int}}
    \ \cocoPar\ 
    \cocoFreeze z {\pmv A_3 \cocoSays \PRECEIVE{b}{bool}}
  }
  \ldots
  \]
  The $\cocoFuse$ prefix cannot be fired: no contract matches $\ptpc
  A_3$'s, and thus the three latent contracts cannot be assigned a
  global type.
  However, by rearranging the system with congruence $\equiv$,
  we have:
  \[
  \ldots
  \cocoSys {\ptpc{B}} {\cocoFuse \cocoSeq P \cocoPar Q}
  \ \ \cocoPar\ \ 
  \cocoSys {\ptpc{B}} {
    \cocoFreeze x {\pmv A_1 \cocoSays \PSEND{a}{int}}
    \ \cocoPar\ 
    \cocoFreeze y {\pmv A_2 \cocoSays \PRECEIVE{a'}{int}}
  }
  \ \ \cocoPar\ \ 
  \cocoSys {\ptpc{B}} {
    \cocoFreeze z {\pmv A_3 \cocoSays \PRECEIVE{b}{bool}}
  }
  \ldots
  \]
  It is now possible to synthesise a global type
  $\GSEND{A_1}{A_2}{int}$, and a session may be created for $\ptpc
  A_1$ and $\ptpc A_2$.  $\ptpc A_3$'s latent contract may be
  $\cocoFuse$d later on.
\end{example}

\input{flexibility.tex}

\input{honesty.tex}


%% file: flexibility.tex
\subsection{Flexibility of Session Establishment}
\label{sec:flexiblity}
We highlight the flexibility of our definition of contract agreement,
together with the semantics of \coco, by discussing three features:
($i$) contracts may use a mix of participant names and variables,
($ii$) different contracts may use common participant variables, and
($iii$) the definition of agreement may be easily extended.
%

%
\paragraph{Contracts with both participant names and variables.}
$\ptpc{A}$ may want to sell
an item to a \emph{specific} participant $\ptpc{B}$, via \emph{any}
shipping company that provides a package tracking system.
$\ptpc A$'s contract may be:
\[
\PSEND{\ptpc{B}}{price} \PSEP
\PRECEIVE{\ptpc{B}}{ack} \PSEP
\PSEND{\ptpv a}{request} \PSEP
\PRECEIVE{\ptpv a}{tracking} \PSEP
\PSEND{\ptpc B}{tracking}
\]
saying that the seller $\ptpc{A}$ must send a $\msgsort{price}$ to the
buyer $\ptpc B$; once $\ptpc{B}$ has $\msgsort{ack}$nowledged, $\ptpc{A}$
must send a shipping
$\msgsort{request}$ to a shipper $\ptpv a$ --- who must send back a
$\msgsort{tracking}$ number, which is then forwarded to $\ptpc{B}$.
This contract may be fused only if $\ptpc B$ takes part in the session,
while the role of shipper $\ptpv a$
may be played by any participant.
%

\paragraph{Contracts sharing participant variables.}
Consider the \coco process:
\[
\ldots
\cocoSys {\pmv A} {
  \cocoTell{A\!\!\!}{x\!\!\!}{(\PSEND{\ptpv b}{request})}
  \ldots
  X(\vec z, \ptpv{b})
  }\ldots
  \quad
  \text{where $X(\vec z , \ptpv{b}) \defeq
    {(y, \ptpv{b'})
      \cocoTell{A\!\!\!}{y\!\!\!}{(
        \PSEND{\ptpv b'}{quote} \PSEP
        \ldots
        \PSEND{\ptpv b}{address}
        )}
      \ldots
      X(\vec z , \ptpv{b})
    }$}
\]
Here, $\ptpc A$ advertises two contracts: the first one ($\PSEND{\ptpv
  b}{request}$) is used by $\ptpc A$ to find a shipping company, and
the second ($ \PSEND{\ptpv b'}{quote} \PSEP \ldots \PSEND{\ptpv
  b}{address}$) to sell items.
The two contracts are linked by the common variable $\ptpv{b}$: whenever
the first one is fused,
variable $\ptpv b$ is instantiated to a participant name, say $\ptpc
B$, which is also substituted in the second. This means that
whenever a new selling session starts, $\ptpc B$ will also be
involved as the receiver of the $\msgsort{address}$ message.




\input extension




%% file: extension.tex
\paragraph{Possible extensions.}
The participants firing $\cocoFuse$-primitives are playing the role
of brokers in our framework.
Depending on their implementation, brokers may also have some
obligations in the contracts they $\cocoFuse$,
or they may want to enforce some general policy
--- therefore they may have additional requirements before agreeing
to start a session.
%
For instance, a broker may not want to start a session with too many
participants as it may be too resource demanding (too many connections etc.).
Another broker may want to start sessions that terminates after a limited
number of interactions, because it has a short life expectancy, e.g., due to an approaching 
scheduled maintenance.
Another kind of broker may precisely want to start sessions which do not
terminate, e.g., if the broker is interested in resilient services.

Several variations of the $\cocoFuse$ primitive are possible
thanks to the fact that we base contract agreements on objects representing
the overall choreography.
We introduce
$\cocoFuseN{n}$, a version of $\cocoFuse$ that only fuses
sessions where there are at least $n$ participants;
$\cocoFuseT$, which has the additional constraint that no recursive behaviour is allowed
in the synthesised choreography (therefore ensuring that
the session will eventually terminate),
and $\cocoFuseR$, which only creates sessions when the synthesised
choreography never terminates (i.e., it only consists of recursive behaviours).

The three extensions may be defined directly via small
modifications of Definition~\ref{def:co2:agreement}:
\begin{itemize}
\item $\cocoFuseN{n}$:
  we add the condition $\lvert \PARTS{\GA} \rvert \geq n$,
  where $\PARTS{\GA}$ is the set of participants in $\GA$;
\item $\cocoFuseT$:
  we add the condition that there should not be any
  recursion variable $\GRECV$ in $\GA$;
\item  $\cocoFuseR$:
  we add the condition that $\GEND$ does not appear in $\GA$.
\end{itemize}
This kind of properties must be checked for at global level because
it cannot always be decided by looking at the individual contracts.
For instance, a participant might exhibit a recursive behaviour
in one of the branches of an external choice, while the participant
it interacts with may always choose a branch that is not recursive.
Note that none of these variations actually affect the results that
follow, since the original $\cocoFuse$ primitive is also blocking.
The variations only restrict some of its applications.
Further variations of $\cocoFuse$ are sketched in Section~\ref{sec:conc}, as
future work.


%% file: honesty.tex
\section{The Problem of Honesty} \label{sec:honesty}

In this section, we discuss and define the notion of
\emph{honesty}~\cite{BTZ12}, i.e., the ability of a participant
to always fulfil its contracts, in any context.
Broadly speaking, in our contract-oriented setting, honesty is
the counterpart of well-typedness in a session type setting:
the static proof that a participant always honours its contracts provides
guarantees about its runtime behaviour.

%

As seen in Example~\ref{ex:co2:do}, each $\cocoDo{}{}{}\!$ prefix
within the process of a participant, say $\cocoSys{\ptpc{A}}{P}$,
is driven by the contract that $\ptpc{A}$ promised to abide by.
%
In a sense, \coco is \emph{culpability-driven},
according to Definition~\ref{def:culpability} below: when a
participant is culpable, it has the duty of making the session
progress according to its contract.

\begin{definition}[Culpability] \label{def:culpability} Let $S$ be a
  \coco system with a session $s$, i.e., \( S \equiv (\vec{u},
  \vec{\ptpv{a}}) \left( \cocoSys {\pmv A} P \cocoPar \cocoSys {\sesc
      s} {\clsy{T}} \cocoPar S_1 \right) \).
  We say that $\ptpc{A}$ is \emph{culpable in $S$ (at session $s$)}
  when there exist
  $\ptpc{B}$ and $\atom{e}$ such that $T \cocoCmove{A}{e}{B}$.
\end{definition}

A culpable participant can overcome its status by firing its
$\cocoDo{}{}{}\!$ prefixes, according to $\rulename{\coco-Do}$, until
another participant becomes culpable or session $s$ terminates.
%
Hence, as long as a culpable participant $\ptpc{A}$ does not enable a
$\cocoDo{}{}{}\!$-prefix matching a contractual action,
$\ptpc{A}$ will remain culpable.
Note that when a participant is involved in multiple sessions, it
may result culpable in more than one of them.

When a participant ${\ptpc A}$ is always able to fulfil its
contractual actions (i.e., overcome its culpability), no
matter what other participants do, then it is said to be \emph{honest}
(cf.\ Definition~\ref{def:honest-readiness}).
This is a desirable property in a distributed contract-oriented
scenario: 
a participant may be stuck in a culpable condition either due to
``simple'' bugs (cf.\ Example~\ref{ex:honesty:readydo-weakreadydo}),
or due to the unexpected (or malicious) behaviour of other participants
(cf.\ Example~\ref{ex:result:multiple-sessions}).
Therefore, before deploying a service, its developers might want to
ensure that it will always be able exculpate itself.

Formally, as in \cite{BSTZ13typing}, we base the definition of honesty
on the relationship between the ready sets of a contract,
and those of a \coco process.
We call the former \emph{contract ready sets}, and the latter
\emph{process ready sets}.
The concept of contract ready sets is similar to
\cite{Castagna09toplas,BTZ12,BSTZ13typing}, where only bilateral
contracts are considered.
Here, we adapt it to suit our multiparty contract model.
%
\begin{definition}[Contract Ready Sets] \label{def:rs}
  The \emph{ready sets of a contract $\PV$},
  written $\cocoAllReadysets{\PV}$, are:
  \[
  \cocoAllReadysets{\PV} = \begin{cases}
    \cocoAllReadysets{\clc{\PV'}} & \text{if\ \ $\PV = \PRECN \clc{\PV'}$}
    \\
    \setcomp{\setenum{(\ptpc{A}_i, \atom{e}_i)}}{i \in I}
    & \text{if\ \ $\PV = \PINS{i \in I}{\PSEND{A_i}{\atom{e}_i}{\PSEP c_i}}$
      \ \ and \ \ $I \neq \emptyset$}
    \\
    \setenum{\setcomp{(\ptpc{A}, \atom{e}_i)}{i \in I}}
    & \text{if\ \ $\PV = \PSUM{i \in I}{\PRECEIVE{A}{\atom{e}_i}{\PSEP c_i}}$}
  \end{cases}
  \]
  %
\end{definition}
Intuitively, when a participant $\ptpc{A}$ is bound to a contract $\PV$,
the ready sets of $\PV$ tell which interactions $\ptpc{A}$ must be able
to perform towards other participants.
Each interaction has the form of a pair, consisting of a participant name
and a message sort.
The interactions offered by an external choice are all available at once,
while those offered by an internal choice are mutually exclusive.

\def\cboneSubstReductChoiceRS{
  \setenum{
    \setenum{(\ptpc{B}_2, \atom{ok})}, \setenum{(\ptpc{B}_2,
      \atom{bye})}}
}
\begin{example} \label{ex:ready-sets-store-customers} Consider the
  system of contracts $\clsy{\SYSV_{\ptpc{A}\ptpc{B}_1\ptpc{B}_2}}$
  from Example~\ref{ex:session:judgement} --- and in particular, the
  stipulated contracts therein, with substitution $\pi = \piSubst$
  from Example~\ref{ex:co2:agreement}:
  \[
  \begin{array}{lcccl}
    \clc{\tilde{\PV_{\ptpc A}}} &=& \clc{\PV_{\ptpc{A}} \pi} &=& \caSubst \\[.1pc]
    \clc{\tilde{\PV_{\ptpc{B}_1}}} &=& \clc{\PV_{\ptpc{B}_1} \pi} &=& \cboneSubst\\[.1pc]
    \clc{\tilde{\PV_{\ptpc{B}_2}}} &=& \clc{\PV_{\ptpc{B}_2} \pi} &=& \cbtwoSubst\\[.1pc]
  \end{array}
  \]
  We have $\cocoAllReadysets{\clc{\tilde{c_{\ptpc A}}}} =
  \setenum{\setenum{(\ptpc{B}_1, \atom{req})}}$:
  in other words, at this
  point of the contract, an interaction is expected between $\ptpc{A}$
  and $\ptpc{B}_1$ (since $\ptpc{A}$ is waiting for $\atom{req}$),
  while no interaction is expected between $\ptpc{A}$ and
  $\ptpc{B}_2$.

  Let us now equip $\clsy{\SYSV_{\ptpc{A}\ptpc{B}_1\ptpc{B}_2}}$ with
  one queue between each pair of participants, and let it perform the
  $\atom{req}$uest exchange between $\ptpc{B}_1$ and $\ptpc{A}$, with
  the transitions:
  \[
  \clsy{\SYSV_{\ptpc{A}\ptpc{B}_1\ptpc{B}_2}}
  \PPAR \QUEUES{\SYSV_{\ptpc{A}\ptpc{B}_1\ptpc{B}_2}}
  \semSend{\ptpc{B}_1}{req}{\ptpc{A}}
  \semRecv{\ptpc{A}}{req}{\ptpc{B}_1}
  \clsy{\SYSV_{\ptpc{A}\ptpc{B}_1\ptpc{B}_2}'} \PPAR
  \QUEUES{\SYSV_{\ptpc{A}\ptpc{B}_1\ptpc{B}_2}}
  \]
  We have that $\clc{\tilde{\PV_{\ptpc A}}}$ in
  $\clsy{\SYSV_{\ptpc{A}\ptpc{B}_1\ptpc{B}_2}'}$ is now reduced to:
  \[
  \clc{\tilde{\PV_{\ptpc A}}'} = \caSubstReduct
  \]
  and thus we have $\cocoAllReadysets{\clc{\tilde{\PV_{\ptpc A}}'}} =
  \setenum{\setenum{(\ptpc{B}_2, \atom{req})}}$, i.e.,
  $\ptpc{A}$ is
  now waiting for a $\atom{req}$uest from $\ptpc{B}_2$.

  %
  If we let the system reduce further, $\clc{\tilde{\PV_{\ptpc A}}'}$ reaches
  its external choice:
  \[
  \clc{\tilde{\PV_{\ptpc A}}''} = \caSubstReductChoice
  \]
  Now, the ready sets become $\cocoAllReadysets{\clc{\tilde{\PV_{\ptpc
          A}}''}} = \setenum{ \setenum{(\ptpc{B}_1, \atom{order}),
      (\ptpc{B}_1, \atom{bye})}}$, i.e., $\ptpc{A}$ must handle both
  answers from $\ptpc{B}_1$.
  Instead, when $\clc{\tilde{\PV_{\ptpc{B}_1}}}$ reduces to its
  internal choice, we have:
  \[
  \clc{\tilde{\PV_{\ptpc{B}_1}}''} = \cboneSubstReductChoice
  \]
  Thus, its ready sets become
  $\cocoAllReadysets{\clc{\tilde{\PV_{\ptpc{B}_1}}''}} =
  \cboneSubstReductChoiceRS$: $\ptpc{B}_1$ is free to choose
  either branch.
\end{example}


Example~\ref{ex:ready-sets-store-customers} shows that, when a
\emph{contract} $\PV$ of a principal $\ptpc{A}$ evolves within a system $\SYSV$,
its ready sets change.
Now we need to define the counterpart of contract ready sets for
\coco processes, i.e., the \emph{process ready sets}.
Again, we adapt the definition from \cite{BSTZ13typing} to our
multiparty contract model.

\begin{definition}[Process Ready Set] \label{def:co2:readydo} For all
  \coco systems $S$, all participants $\ptpc{A}, \ptpc{B}$ and
  sessions $u$, we define the set of pairs:\footnote{The side contition
    ``$u \not\in\vec {v}$" of Definition~\ref{def:co2:readydo} deals with
    cases like
    \(
    S_0 = (s)\left(\cocoSys{\ptpc{A}}{\cocoDo{s}{\ptpc{B}}{int}}\right)
    \)
    and
    \(
    S = S_0 \cocoPar \cocoSys{s}{\PBOX{A}{\PSEND{B}{int}} \PPAR \clsy{\ldots}}
    \cocoPar \ldots
    \):
    without the side condition, $\cocoAllReadydo{u}{A}{S_0} = \setenum{
      \setenum{(\ptpc{B}, \atom{int})}}$ --- hence, by
    Def.~\ref{def:readiness}, $\ptpc{A}$ would result to be ready in $S$.}
  \[
  \cocoAllReadydo{u}{A}{S} =
  \setcomp{(\ptpc{B}, \atom{e})\ }
  {\ \exists \, \vec{v}, \vec{\ptpv{a}}, P,P',Q,S' \;\suchthat\; {
      S \equiv (\vec{v}, \vec{\ptpv{a}}) \left({
          \cocoSys {\ptpc{A}} {\cocoDo{u}{\ptpc B}{e}
            \cocoSeq P \cocoPlus P' \cocoPar Q} \cocoPar S_0
        }\right) \cocoPar S_1
      \;\;\land\;\; u \not \in \vec{v}
    }}
  \]
\end{definition}

Intuitively, Definition~\ref{def:co2:readydo} says that the process ready
set of $\ptpc{A}$ over a session $u$ in a system $S$ contains the
interactions that $\ptpc{A}$ is immediately able to perform with other
participants through its $\cocoDo{u}{\_}{}$ prefixes.
Just as in contract ready sets, the interactions are represented by
participant/sort pairs.

Next, we want to characterise a weaker notion of the process ready set,
so it only takes into account the first actions \emph{on a specific session}
that a participant is ready to make.

\begin{definition}[Weak Process Ready Set] \label{def:co2:readydo-weak}
  We write $S \cocoPassivearrow{u}{A}{} S'$ iff:
  \[
  \exists \, \ptpc{B} , p
  \suchthat
  S \xrightarrow{{\ptpc{B}} \colon p} S'
  \;  \implies \;
  \left(
    \ptpc{A} \neq \ptpc{B}
    \quad \lor \quad
    \forall \, \SORTV \suchthat
    \forall \, \ptpc{C}
    \suchthat
    p =  \cocoDo{v}{\ptpc{C}}{\atom \SORTV}
    \implies
    u \neq v 
  \right)
  \]

  We then define the set of pairs
  $\cocoAllWeakreadydo{u}{A}{S}$ as:
  \[
  \cocoAllWeakreadydo{u}{A}{S}
  \; = \;
  \setcomp{(\ptpc{B}, \atom{e})\ }{\ 
    \exists S' \suchthat S \cocoPassivearrow{u}{A}{B}{\!\!}^* \; S'
    \;\land\; (\ptpc{B}, \atom{e}) \in \cocoAllReadydo{u}{A}{S'}}
  \]
\end{definition}

In Definition~\ref{def:co2:readydo-weak}, we are not interested
in the actions that do not relate to the session $u$. Thus, we allow
the system to evolve either by
($i$) letting any other participant other than $\ptpc{A}$ do an action,
or ($ii$) letting $\ptpc{A}$ act on a different session than $u$,
or ($iii$) do internal actions.

We now introduce the final ingredient for honesty, that is the notion of
\emph{readiness} of a participant.

\begin{definition}[Readiness] \label{def:readiness}
  We say that ${\ptpc A}$ is ready in $S$ iff, whenever $S \equiv
  (\vec{u}, \vec{\ptpv b}) S_0$ for some $\vec{u},
  \vec{\ptpv b}$ and $S_0 \equiv {\cocoSys s {\PBOX{A}{c} \PPAR \cdots}}
  \cocoPar \cdots$, the following holds:
  \[
  \exists \mathcal{X} \in \cocoAllReadysets{\PV} \suchthat
  \left( (\ptpc{B}, \atom{e}) \in \mathcal{X}
    \implies
    (\ptpc{B}, \atom{e}) \in \cocoAllWeakreadydo{s}{A}{S_1}
  \right)
  \]
\end{definition}

Definition~\ref{def:readiness} says that a participant
$\ptpc{A}$ is $\emph{ready}$ in a system $S$ whenever its process ready
sets for a session $s$ will eventually
contain all the participant/sort pairs of one of the
contract ready sets of $\ptpc{A}$'s contract in $s$.
When a participant $\ptpc{A}$ is ``ready'', then, for any of its
contracts $\PV$, the \coco process of $\ptpc{A}$ is (eventually)
able to fulfil at least the interactions in $\PV$'s prefix.

\def\paReductChoice{
  {\cocoDo{x}{\ptpv{b_1}}{order} \cocoSeq \cocoDo{x}{\ptpv{b_2}}{ok}}
  \;+\;
  {\cocoDo{x}{\ptpv{b_1}}{bye} \cocoSeq \cocoDo{x}{\ptpv{b_2}}{bye}}
}
\def\pboneReduct{
  \cocoDo{y}{\ptpv{a}}{order}
}
\def\paSubstReductChoice{
  {\cocoDo{s}{\ptpc{B}_1}{order} \cocoSeq \cocoDo{s}{\ptpc{B}_2}{ok}}
  \;+\;
  {\cocoDo{s}{\ptpc{B}_1}{bye} \cocoSeq \cocoDo{s}{\ptpc{B}_2}{bye}}
}
\def\pboneSubstReduct{
  \cocoDo{s}{\ptpc{A}}{order}
}
\def\pbtwoSubstReductChoice{
  {\cocoDo{s}{\ptpc{B}_1}{ok} \cocoSeq \cocoDo{s}{\ptpc{A}}{ok}}
  \;+\;
  {\cocoDo{s}{\ptpc{B}_1}{bye} \cocoSeq \cocoDo{s}{\ptpc{A}}{bye}}
}
\begin{example} \label{ex:honesty:readydo-weakreadydo}
  We have seen that, after fusion of the latent contracts of $S_1$ (in
  Ex.~\ref{ex:co2:agreement}) we obtain:
  \[
  (s)S_1' \ \ \ \equiv\ \ \ {(s)\big(
    {\cocoSys{\ptpc{A}}{P_{\ptpc{A}}\sigma\pi}}
    \ \ \cocoPar\ \ 
    {\cocoSys{s}{
        \clsy{
          \SYSV_{\ptpc{A}\ptpc{B}_1\ptpc{B}_2}
          \cocoPar \QUEUES{\SYSV_{\ptpc{A}\ptpc{B}_1\ptpc{B}_2}}}
      }
    }
    \ \ \cocoPar\ \ 
    {\cocoSys{\ptpc{B}_{1}}{P_{\ptpc{B}_{1}}\sigma\pi}}
    \ \ \cocoPar\ \ 
    {\cocoSys{\ptpc{B}_{2}}{P_{\ptpc{B}_{2}}\sigma\pi}}
    \big)}
  \]
  where the substitutions $\sigma$ and $\pi$ are also from
  Ex.~\ref{ex:co2:agreement}.
  Let us define the processes (after substitutions):
  \[
  \begin{array}{lcl}
    P_{\ptpc{A}}\sigma\pi &\;=\;& {\cocoDo{s}{\ptpc{B}_1}{req}
      \cocoSeq \cocoDo{s}{\ptpc{B}_2}{req}
      \cocoSeq \cocoDo{s}{\ptpc{B}_1}{quote}
      \cocoSeq \big({
        \paSubstReductChoice
      }\big)
    } \\
    P_{\ptpc{B}_1}\sigma\pi &\;=\;& {\cocoTau \cocoSeq \cocoDo{s}{\ptpc{A}}{req}
      \cocoSeq \cocoDo{s}{\ptpc{A}}{quote}
      \cocoSeq \pboneSubstReduct
    } \\
    P_{\ptpc{B}_2}\sigma\pi &\;=\;& {\cocoDo{s}{\ptpc{A}}{req}
      \cocoSeq \big(
      \pbtwoSubstReductChoice
      \big)
    }
  \end{array}
  \]

  Thus, we have:
  \[
  \begin{array}{lcccl}
    \cocoAllReadydo{s}{A}{S_1'} &=& \setenum{(\ptpc{B}_1, \atom{req})}
    &=& \cocoAllWeakreadydo{s}{A}{S_1'} \\
    \cocoAllReadydo{s}{B_1}{S_1'} &=& \emptyset
    \;\neq\; \setenum{(\ptpc{A}, \atom{req})}
    &=& \cocoAllWeakreadydo{s}{B_1}{S_1'} \\
    \cocoAllReadydo{s}{B_2}{S_1'} &=& \setenum{(\ptpc{A}, \atom{req})}
    &=& \cocoAllWeakreadydo{s}{B_2}{S_1'}
  \end{array}
  \]
  Note that the $\cocoTau$ prefix in $P_{\ptpc{B}_1}$ prevents
  $\ptpc{B}_1$ from interacting immediately with $\ptpc{A}$
  on session $s$, although it is ``weakly ready'' to do it.
  Hence, considering that the weak process ready sets of each participant
  in $S_1'$ match their respective contract ready sets in
  $\clsy{\SYSV_{\ptpc{A}\ptpc{B}_1\ptpc{B}_2}}$
  (Example~\ref{ex:ready-sets-store-customers}) according to
  Definition~\ref{def:readiness} we have that participants
  $\ptpc{A}$, $\ptpc{B}_1$ and $\ptpc{B}_2$ are all ready in $(s)S_1'$.
\end{example}

Before defining honesty formally, we need to characterise the class
of systems for which this concept is meaningful, 
i.e., those systems where a participant
is not (yet) involved in latent contracts nor active sessions.
\begin{definition}[Initial System]
  A \coco system $\CSY$ is \emph{$\pmv{A}$-initial} if $\CSY$ has no
  sub-term of the form $\cocoFreeze \_ {\pmv A \cocoSays \PV}$ or
  $\clsy{\PBOX{\ptpc{A}}{\PV}}$
  with $\PV \not\equiv \PEND $.
  A \coco system $\CSY$ is \emph{initial} when it is $\pmv{A}$-initial
  for each participant $\ptpc{A}$ in $\CSY$.
\end{definition}

%
\begin{definition}[Honesty] \label{def:honest-readiness} We say that
  $\cocoSys{\ptpc A}{P}$ is \emph{honest} iff, for all $\pmv{A}$-initial
  $S \equiv {
    (\vec{u}, \vec{\ptpv{b}})\left(\cocoSys{\pmv{A}}{P} \cocoPar S_0\right)}$
  s.t.\ $S \rightarrow^* S'$,
  ${\ptpc A}$ is ready in $S'$.
\end{definition}

A process $\cocoSys{\ptpc A}{P}$
is said to be honest when, for all contexts and reductions that
$\cocoSys{\ptpc A}{P}$ may be engaged in, ${\ptpc A}$ is persistently
ready.
In other words, there is a continuous correspondence
between the interactions exposed in the contract ready sets
and the process ready sets of the possible reductions of
any system involving $\cocoSys{\ptpc A}{P}$.
The definition rules out contexts with latent/stipulated contracts of
$\ptpc{A}$, otherwise $\ptpc{A}$ could be made trivially
dishonest, e.g.~by inserting a latent contract $\cocoFreeze u
{\ptpc{A} \cocoSays \PV}$ that $\ptpc{A}$ cannot fulfil.

\def\exDishonestyAllEmptyQueues{
  \QUEUE{A B_1}{\emptyQ} \PPAR \QUEUE{B_1 A}{\emptyQ}
  \PPAR \QUEUE{A B_2}{\emptyQ} \PPAR \QUEUE{B_2 A}{\emptyQ}
  \PPAR \QUEUE{B_1 B_2}{\emptyQ} \PPAR \QUEUE{B_2 B_1}{\emptyQ}
}
\begin{example} \label{ex:honesty:dishonesty}
  Consider the process
  $\cocoSys{\ptpc{B}_{1}}{\cocoTell{\ptpc{A}}{y}{c_{\ptpc{B}_{1}}}
    \cocoSeq P_{\ptpc{B}_{1}}}$ of system $S_1$, as defined in
  Examples~\ref{ex:co2:agreement} and \ref{ex:honesty:readydo-weakreadydo}.
  We show that this process is \emph{not} honest.
  In fact, $S_1$ can reduce as $S_1 \cocoMove{}^* (s)S'_1
  \cocoMove{}^* (s)S_1''$, where:
  \begin{IEEEeqnarray*}{rclll}
    (s)S_1'' &\;\;=\;\;&
    (s)\Big(\;&\IEEEeqnarraymulticol{2}{l}{
      {\cocoSys{\ptpc{A}}{\paSubstReductChoice}}} \\
    & & &\ \cocoPar\ \ 
    s\big[\;& 
    \clsy{
      \PBOX{\ptpc{A}}{\caSubstReductChoice}} \\
    & & & & \clsy{\;\PPAR\;
      \PBOX{\ptpc{B}_1}{\cboneSubstReductChoice} \;\PPAR\;
      \PBOX{\ptpc{B}_2}{\cbtwoSubstReductChoice}} \\
    & & & & {
      \;\PPAR \exDishonestyAllEmptyQueues
    } \;\big] \\
    & & & \IEEEeqnarraymulticol{2}{l}{
      \ \cocoPar\ \ 
      {{\cocoSys{\ptpc{B}_{1}}{\pboneSubstReduct}}
        \ \ \cocoPar\ \ 
        {\cocoSys{\ptpc{B}_{2}}{\pbtwoSubstReductChoice}}
        \Big)
      }
    }
  \end{IEEEeqnarray*}
  At this point, we see that there is a problem in the
  implementation of $\ptpc{B}_1$:
  it does not notify the other buyer before making an order.
  In fact, $\ptpc{B}_1$'s process is trying to perform
  $\pboneSubstReduct$, but its contract requires that
  $\cocoDo{s}{\ptpc{B}_2}{ok}$ is performed first (or
  $\cocoDo{s}{\ptpc{B}_2}{bye}$, if the quote is rejected).
  This is reflected by the mismatch between $\ptpc{B}_1$'s process ready set in
  $S_1''$, and its contract ready sets, in session $s$:
  \[
  \begin{array}{rcl}
    \cocoAllReadydo{s}{B_1}{S_1''} &=& \setenum{\setenum{
        (\ptpc{A}, \atom{order})}} \\
    \cocoAllReadysets{\cboneSubstReductChoice} &=& \cboneSubstReductChoiceRS
  \end{array}
  \]
  In terms of the above definitions,
  there exists a system $S_1$ --- containing
  $\cocoSys{\ptpc{B}_{1}}{\cocoTell{\ptpc{A}}{y}{c_{\ptpc{B}_{1}}}
    \cocoSeq P_{\ptpc{B}_{1}}}$ --- that reduces
  to a $(s)S_1''$ where $\ptpc{B}_1$ is not ready
  (Definition~\ref{def:readiness}).
  Therefore,
  $\cocoSys{\ptpc{B}_{1}}{\cocoTell{\ptpc{A}}{y}{c_{\ptpc{B}_{1}}}
    \cocoSeq P_{\ptpc{B}_{1}}}$ is not honest.
  In fact, $\ptpc{B}_1$ is culpable in $(s)S_1''$, according to
  Definition~\ref{def:culpability}.
\end{example}

As in \cite{BSTZ13typing}, the definition of honesty subsumes a
\emph{fair} scheduler, eventually allowing participants to fire
persistently (weakly) enabled $\cocoDo{}{}{\!\!}$ actions.

Honesty is not decidable in general~\cite{BTZ12},
but for a bilateral contract model it has been approximated either 
via an abstract
semantics~\cite{BTZ12} or a type discipline~\cite{BSTZ13typing} for
\coco. 
We believe that these approximations may be easily adapted to our
setting (see Section \ref{sec:conc} for more details).



%% file: result.tex
\section{Results} \label{sec:result}

We now give the main properties of our framework.
We ensure that two basic features of \coco hold in our multiparty
adaptation: the state of a session always allows to establish who is
responsible for making the system progress
(Th.~\ref{thm:culpability}) and honest participants can always
exculpate themselves (Th.~\ref{thm:exculpation}).
We then formalise a link between the honesty of participants, and two
key properties borrowed from the session types setting:
Th.~\ref{thm:fidelity} introduces session fidelity in \coco;
and Th.~\ref{thm:progress} introduces a notion of progress in \coco,
based on the progress of the contractual agreement (and its choreography).

\begin{theorem}[Unambiguous Culpability]\label{thm:culpability}
  Given an initial \coco system $S$, if $S \rightarrow^* S' \equiv
  (\vec{u},\vec{b})\left(\cocoSys{s}{\SYSV} \cocoPar \cdots \right)$
  such that $\SYSV \not\equiv \END$, then there exists
  at least one culpable participant in $S'$.
\end{theorem}
Theorem~\ref{thm:culpability} says that in an active session established
through a $\cocoFuse$ reduction, there is
always at least one participant $\cocoSys{\ptpc{A}}{P}$ who leads the
next interaction.
Thus, if a corresponding $\cocoDo{s}{\ptpc{B}}{\!e}$ prefix is not in
$P$, $S$ may get stuck, and $\ptpc{A}$ is culpable.

\begin{example} \label{ex:results:culpability}
  Consider the system $S_1''$ in Example~\ref{ex:honesty:dishonesty}, and
  the system of contracts in its session $s$:
  \[
  \begin{array}{rcl}
    \clsy{T_s} &\;=\;& \clsy{\PBOX{\ptpc{A}}{\caSubstReductChoice}} \\[.2pc]
    & & \;\clsy{\PPAR\;
      \PBOX{\ptpc{B}_1}{\cboneSubstReductChoice} \;\PPAR\;
      \PBOX{\ptpc{B}_2}{\cbtwoSubstReductChoice}} \\[.2pc]
    & & \;\PPAR\; \exDishonestyAllEmptyQueues
  \end{array}
  \]
  We have $\clsy{T_s} \cocoCmove{B_1}{ok}{B_2}$ and $\clsy{T_s}
  \cocoCmove{B_1}{bye}{B_2}$.
  Hence, $\ptpc{B}_1$ is responsible for the next interaction, and
  culpable for $S_1''$ being stuck.
\end{example}

\begin{theorem}[Exculpation] \label{thm:exculpation}
  Given an $\pmv{A}$-initial \coco system $S_0$ with
  $\cocoSys{\ptpc{A}}{P}$ honest, whenever $S_0 \cocoMove{}^* S \equiv
  (\vec{u}, \vec{\ptpv{a}}) \left( \cocoSys {\sesc s} \SYSV \cocoPar
    S_1 \right)$ and $\ptpc{A}$ is culpable in $S$ at session $s$, there
  exist $\ptpc{B}$ and $\msgsort{e}$ such that:
  $
  S \cocoMove{\ptpc{A}\colon
    p}{}\!\!\!^*\cocoMove{\ptpc{A}\colon
    \cocoDo{\ptpc{B}}{s}{e}}
  $
  where $p = \cocoTau$ or $p = \cocoTell{\_}{\_}{\_}$.
\end{theorem}

Theorem~\ref{thm:exculpation} follows from the definition of honesty,
formalising that honest participants can always overcome
their culpability, by firing their contractual
$\cocoDo{}{}{}\!$ actions (possibly after advertising other contracts
or performing some internal actions).

\begin{theorem}[Fidelity]\label{thm:fidelity}
  For all initial systems $S$ with only honest participants,
  if $S$ is
  such that
  $S
  \cocoMove{}^* S' \equiv  
  (\vec{u}, \vec{\ptpv{a}})
  \left( \cocoSys
    {\pmv A} P \cocoPar
    \cocoSys {\sesc s} \SYSV \cocoPar S_0 \right)$, then
  \(
  (
  S'
  \cocoPassivearrow{s}{A}{}{}\!\!\!^* \cocoMove{{\ptpc A}
    \colon \cocoDo{s}{\ptpc B}{e} }
  )
  \iff
  (
  \SYSV 
  \cocoCmove{A}{e}{B}
  )
  \)
  \ \ (where $\cocoPassivearrow{u}{A}{}\!\!\!^*$, as in
  Def.~\ref{def:co2:readydo-weak}, intuitively stands for any reduction
  \emph{not} involving session $s$).
\end{theorem}

Theorem~\ref{thm:fidelity} says that each (honest)
participant will strictly adhere to its contracts, once they have
been fused in a session.
It follows directly from the semantics of \coco (that forbid
non-contractual $\cocoDo{}{}{\!}$ prefixes to be fired) and from the
definition of honesty.

Theorem~\ref{thm:progress} below introduces the notion of global progress,
which is slightly different from the contractual progress.
In fact, progress in \coco is only meaningful \emph{after} a session
has been established, and thus a culpable participant exists.
A system without sessions may not progress because a set of
compliant contracts cannot be found, or a $\cocoFuse$ prefix is not
enabled.
In both cases, no participant may be deemed culpable, and thus
responsible for the next move.
However, the system may progress again if other (honest) participants
join it, allowing a session to be established.

\begin{theorem}[Global Progress]\label{thm:progress}
  Given an initial \coco system $S_0$ with only honest participants,
  if
  $S_0 \cocoMove{}^* S \equiv (\vec{u}, \vec{\ptpv{a}}) \left(
    \cocoSys {\sesc s} \SYSV \cocoPar S_1 \right)$ with
  $\SYSV \not\equiv \END$, then $S \cocoMove{}$.
\end{theorem}

Theorem~\ref{thm:progress} follows from the definition of honesty
(i.e., participants are always ready to fulfil their contracts), the
fact that contract compliance guarantees contractual progress
\cite{LT12}, Theorem~\ref{thm:exculpation}, and the semantics of
\coco (in particular, rule \rulename{\coco-Do}).
This result also holds for systems where a process takes part in
multiple sessions: the honesty of all participants guarantees that all
sessions will be completed. 

\begin{example} \label{ex:result:multiple-sessions}
  We now give a simple example on a system with multiple sessions.
  We show how a seemingly honest process ($\ptpc{B}$) could be deemed
  culpable due to the unexpected behaviour of other participants, and
  how honest participants guarantee progress of the whole system.
  %
  %
  %
  Consider:
  \begin{IEEEeqnarray*}{rcll}
    S  &\;=\; & (x,y,z,w) \big(\;&
    {\cocoSys{\ptpc{A}}{
        \cocoTell{\ptpc{A}}{x}{\clc{(\PSEND{B}{int})}}
        \cocoSeq
        \cocoFuse \cocoSeq \cocoFuse
      }
    }
    \ \ \ \cocoPar\ \ \ 
    {\cocoSys{\ptpc{B}}{
        \cocoTell{\ptpc{A}}{y}{\clc{(\PRECEIVE{A}{int})}}
        \cocoSeq
        \cocoTell{\ptpc{A}}{z}{\clc{(\PSEND{C}{bool})}}
        \cocoSeq
        \cocoDo {y} {\ptpc A} {int}
        \cocoSeq
        \cocoDo {z} {\ptpc C} {bool}
      }
    } \\
    & & &
    \ \cocoPar\ \ \ 
    {\cocoSys{\ptpc{C}}{
        \cocoTell{\ptpc{A}}{w}{\clc{(\PRECEIVE{B}{bool})}}
        \cocoSeq
        \cocoDo {w} {\ptpc{B}} {bool}
      }
    }\;\big)
  \end{IEEEeqnarray*}
  After all four contracts have been advertised to $\ptpc{A}$ and
  $\cocoFuse$d, the system reduces to:
  \begin{IEEEeqnarray*}{rcll}
    S'  &\;=\; & (s_1, s_2) \big(\;&
    {\cocoSys{\ptpc{A}}{
        \cocoEmptyproc
      }
    }
    \ \ \ \cocoPar\ \ \ 
    {\cocoSys{\ptpc{B}}{
        \cocoDo {s_1} {\ptpc A} {int}
        \cocoSeq
        \cocoDo {s_2} {\ptpc C} {bool}
      }
    }
    \ \ \ \cocoPar\ \ \ 
    {\cocoSys{\ptpc{C}}{
        \cocoDo {s_2} {\ptpc{B}} {bool}
      }
    } \\
    & & &
    \ \cocoPar\ 
    \cocoSys {s_1} {
      \PBOX{A}{\PRECEIVE{B}{int}}
      \PPAR
      \PBOX{B}{\PSEND{A}{int}}
      \PPAR
      \QUEUE{AB}{\emptyQ} \PPAR \QUEUE{BA}{\emptyQ}
    }
    \ \cocoPar\ 
    \cocoSys {s_2} {
      \PBOX{B}{\PSEND{C}{bool}}
      \PPAR
      \PBOX{C}{\PRECEIVE{B}{bool}}
      \PPAR
      \QUEUE{BC}{\emptyQ} \PPAR \QUEUE{CB}{\emptyQ}
    }\;\big)
  \end{IEEEeqnarray*}
  Even if both sessions $s_1$ and $s_2$ enjoy contractual progress,
  $S'$ is stuck: $\ptpc{A}$ does not perform the promised action, thus
  remaining culpable in $s_1$; $\ptpc{B}$ is stuck waiting 
  in $s_1$, thus remaining culpable in $s_2$.\footnote{In
    this case, $\ptpc{B}$ is deemed culpable in $s_2$ because its
    implementation did not expect $\ptpc{A}$ to misbehave.}
  Indeed, neither $\ptpc{A}$ nor $\ptpc{B}$ are ready in $S'$, and
  thus they are not honest in $S$.
  Hence, global progress is not guaranteed.
  Let us now consider the following variant of $S$, where all participants
  are honest:
  \begin{IEEEeqnarray*}{rcll}
    \hat{S}  &\;=\; & (x,y,z,w) \big(\;&
    {\cocoSys{\ptpc{A}}{
        \left({
            \cocoTell{\ptpc{A}}{x}{\clc{(\PSEND{B}{int})}}
            \cocoSeq
            \cocoDo {x} {\ptpc B} {int}
          }\right)
        \cocoPar
        \cocoFuse
        \cocoPar
        \cocoFuse
      }
    } 
    \quad \cocoPar \quad
    {\cocoSys{\ptpc{C}}{
        \cocoTell{\ptpc{A}}{w}{\clc{(\PRECEIVE{B}{bool})}}
        \cocoSeq
        \cocoDo {w} {\ptpc{B}'} {bool}
      }
    } \\
    & & &
    \ \cocoPar\ \ \ 
    {\cocoSys{\ptpc{B}}{
        \cocoTell{\ptpc{A}}{y}{\clc{(\PRECEIVE{A}{int})}}
        \cocoSeq
        \cocoTell{\ptpc{A}}{z}{\clc{(\PSEND{C}{bool})}}
        \cocoSeq
        \left({
            {\cocoDo {y} {\ptpc A} {int}
              \cocoSeq
              \cocoDo {z} {\ptpc C} {bool}}
            \cocoPlus
            {
              \cocoTau \cocoSeq
              \left({
                  \cocoDo {y} {\ptpc A} {int}
                  \cocoPar
                  \cocoDo {z} {\ptpc C} {bool}
                }\right)
            }
          }\right)
      }
    } \;\big)
  \end{IEEEeqnarray*}
  In this case, $\ptpc{A}$ will respect its contractual duties, while
  $\ptpc{B}$ will be ready to fulfil its contracts on both sessions
  --- even if one is not activated, or remains stuck (here, $\cocoTau$
  represents an internal action, e.g., a timeout: if the first $\cocoDo
  {y} {\ptpc A} {int}$ cannot reduce, $\ptpc{B}$ falls back to running
  the sessions in parallel).
  The honesty of all participants in $\hat{S}$ guarantees that, once
  a session is active, it will reach its completion.
\end{example}



%% file: conc.tex
\section{Concluding remarks}\label{sec:conc}

In this work, we investigated the combination of the
contract-oriented calculus \coco with a
contract model that fulfils two basic design requirements:
$(i)$ it supports multiparty
agreements, and
$(ii)$ it provides an explicit description of the
choreography that embodies each agreement. 
These requirements prompted us towards the well-established results
from the session types setting --- in particular, regarding
the interplay between a well-formed global type and its corresponding
local behaviours.
We introduced the concepts of global progress and session fidelity in \coco,
also inspired from the analogous concepts in theories based on session types.
We built our framework upon a simple version of session types, and yet
it turns out to be quite flexible, allowing for sessions where the
number of participants is not known beforehand.

\textbf{Related work.}
The origin of \coco goes back to~\cite{BZ10}.
The calculus was generalised in \cite{BTZ11ice, BTZ12sacs} to
suit different contract models
(e.g., contracts as processes or logic formul\ae).
In~\cite{BTZ12}, it has been instantiated to a theory of
bilateral contracts inspired by~\cite{Castagna09toplas}.
%
%
A negative result in \cite{BTZ12} is that the problem of honesty
(Section~\ref{sec:honesty}) is not decidable.
A type system for
\coco processes providing a decidable approximation of honesty
was introduced in~\cite{BSTZ13typing}.
This result relies on the product between a finite state system
(approximating contracts) and a Basic Parallel Process
(approximating a \coco process).
%
%
Considering that the systems of contracts in this
work form a (strict) subset of the local/global types
in~\cite{DY12}, for which each configuration is reachable by a
1-buffer execution, we believe that the type system in
\cite{BSTZ13typing} may be adapted to our setting.

The seminal top-down approach of multiparty session types has been first
described in \cite{hyc08}. In summary, the framework works as follows:
designers specify a choreography (i.e., a global type), which is then
projected onto local behaviours (i.e., local types), which in turn
are used to type-check processes.
A dual approach was introduced in~\cite{LT12}: from a set of local types
it is possible to synthesise a choreography (i.e., a global type).
This is precisely the result we use as a basis for our definition of compliant
contracts.

The semantic correspondence between global types and projected local
behaviours has been investigated in \cite{Lanese2008SEFM, cdp11}.
%
%
To the best of our knowledge, no other contract model 
besides ours is based on explicit choreography synthesis.
A related approach is presented in \cite{BravettiZavattaro2008FI}:
multiple contracts are considered compliant when their
composition (i.e., the system of contracts) guarantees completion.
In our work, progress (subsuming completion) is provided
by the synthesis of a global type.
In \cite{BravettiLanese2009LNCS}, contracts are considered
compliant when their composition adheres to a predetermined
choreography;
in our framework, however, no choreography is assumed beforehand.

%
%
%
%

\textbf{Future work.}
We plan to extend our work so to offer even more
flexibility.
For example, by introducing a parameterised $\cocoFuse$ primitive
which starts a session according to different
criteria, when more than one agreement is possible (as in our
introductory example).
For instance, one could choose the agreement involving the most (or least)
number of participants.
These criteria may be based on a semantic characterisation of global
types, e.g., as the ones in~\cite{Lanese2008SEFM,cdp11}.
We also plan to study the possibility for a participant to be
involved in a session under multiple contracts, e.g., a bank
advertising two services, and a customer publishing a contract
which uses both of them 
in a well-formed choreography.

Another research direction is the concept of ``group honesty''.
In fact, the current definition of honesty is quite strict: it basically verifies
each participant in isolation, thus providing a sufficient (but not
necessary) condition for progress.
Consider, for example, a \coco system like:
\[
S \;\;=\;\; (x,y) \big({\cocoSys{\ptpc{A}}{
    \cocoTell{A}{x}{\left(\PSEND{B}{int} \PINCH
        \PSEND{B}{bool}\right)}
    \cocoSeq
    \cocoFuse
    \cocoSeq
    \cocoDo x {\ptpc B} {int}
  }
}
\ \ \cocoPar\ \ 
{\cocoSys{\ptpc{B}}{
    \cocoTell{A}{y}{\left(\PRECEIVE{A}{int} \POUTCH
        \PRECEIVE{A}{bool}\right)}
    \cocoSeq
    \cocoDo y {\ptpc A} {int}
  }
} \big)
\]
$\ptpc{B}$ is dishonest, since it is not ready for the
$\msgsort{bool}$ branch of its contract.
However, the system $S$ has progress: when $\ptpc{B}$ establishes a
session with $\ptpc{A}$, the latter will never take the
$\msgsort{bool}$ branch; hence, $\ptpc{B}$ will not remain culpable.
This kind of ``group honesty'' may be used to
validate (sub-)systems of participants developed by the same organization: it
would ensure that they never ``cheat each other'', and are
collectively honest when deployed in any context.
Furthermore, the group honesty of all participants in a system $S$ may
turn out to be a necessary condition for the global progress of $S$.

\paragraph{Acknowledgements.}
We would like to thank
Massimo Bartoletti, Emilio Tuosto, and Roberto Zunino
for their valuable advice, discussions and comments.
We would also like to thank the anonymous reviewers for their suggestions.
This work is partially supported by 
Aut.\ Region of Sardinia under grants 
L.R.7/2007 CRP-17285 (TRICS) and
P.I.A.\ 2010 project ``Social Glue'',
and by MIUR PRIN 2010-11 project ``Security Horizons'',
and by EU COST Action IC1201
(BETTY).

